\documentclass{IEEEoj}
\usepackage{cite}
\usepackage{amsmath,amssymb,amsfonts}
\usepackage[ruled,vlined]{algorithm2e}
\usepackage{color}
\usepackage{graphicx}
\usepackage{textcomp}
\usepackage{amsmath}
\usepackage{subfigure}

\def\BibTeX{{\rm B\kern-.05em{\sc i\kern-.025em b}\kern-.08em
T\kern-.1667em\lower.7ex\hbox{E}\kern-.125emX}}
\AtBeginDocument{\definecolor{ojcolor}{cmyk}{0.93,0.59,0.15,0.02}}
\def\OJlogo{\vspace{-4pt}\hskip-4pt\includegraphics[height=18pt]{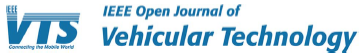}}
\begin{document}

\receiveddate{XX Month, XXXX}
\reviseddate{XX Month, XXXX}
\accepteddate{XX Month, XXXX}
\publisheddate{XX Month, XXXX}
\currentdate{19 December, 2025}
\doiinfo{10.1109/OJVT.2026.3655621}

\title{Delay-Doppler-Domain Channel Estimation and Reduced-Complexity Detection of Faster-than-Nyquist Signaling Aided OTFS}

\author{ZEKUN~HONG\IEEEauthorrefmark{1}, SHINYA~SUGIURA\IEEEauthorrefmark{1},~\IEEEmembership{Senior~Member,~IEEE}, CHAO~XU\IEEEauthorrefmark{2},~\IEEEmembership{Senior~Member,~IEEE}, and LAJOS~HANZO\IEEEauthorrefmark{2},~\IEEEmembership{Life~Fellow,~IEEE}
}
\affil{The Institute of Industrial Science, The University of Tokyo, 153-8505, Japan}
\affil{The School of Electronics and Computer Science, University of Southampton, SO17 1BJ, UK}
\authornote{Preprint (Accepted Version). DOI: 10.1109/OJVT.2026.3655621.
Copyright $\copyright$ 2026 IEEE. Personal use of this material is permitted. However, permission to use this material for any other purposes must be obtained from the IEEE by sending a request to pubs-permissions@ieee.org.
The work of Z. Hong was supported in part by the Japan Science and Technology Agency (JST) SPRING (Grant JPMJSP2108). The work of S. Sugiura was supported in part by JST FOREST (Grant JPMJFR2127), in part by JST ASPIRE (Grant JPMJAP2345), in part by JST CRONOS (Grant JPMJCS25K3), and in part by the Japan Society for the Promotion of Science (JSPS) KAKENHI (Grants 22H01481, 23K22752).}
\markboth{Delay-Doppler-Domain Channel Estimation and Reduced-Complexity Detection of Faster-than-Nyquist Signaling Aided OTFS}{Hong, Sugiura, Xu and Hanzo}

\begin{abstract}
We conceive a novel channel estimation and data detection scheme for OTFS-modulated faster-than-Nyquist (FTN) transmission over doubly selective fading channels, aiming for enhancing the spectral efficiency and Doppler resilience. The delay-Doppler (DD) domain's input-output relationship of OTFS-FTN signaling is derived by employing a root-raised cosine (RRC) shaping filter. More specifically, we design our DD-domain channel estimator for FTN-based pilot transmission, where the pilot symbol interval is lower than that defined by the classic Nyquist criterion. Moreover, we propose a reduced-complexity linear minimum mean square error equalizer, supporting noise whitening, where the FTN-induced inter-symbol interference (ISI) matrix is approximated by a sparse one. 
Our performance results demonstrate that the proposed OTFS-FTN scheme is capable of enhancing the achievable information rate, while attaining a comparable BER performance to both that of its Nyquist-based OTFS counterpart and to other FTN transmission schemes, which employ the same RRC shaping filter.
\end{abstract}


\maketitle

\section{INTRODUCTION}
\IEEEPARstart{C}{lassic} orthogonal frequency-division multiplexing (OFDM) suffers from grave inter-carrier interference in high-mobility environments. To address this issue, orthogonal time frequency space (OTFS) modulation~\cite{hadani2017orthogonal} was conceived for modulation in the delay-Doppler (DD) domain, while leveraging a sparse quasi-static channel representation. In contrast to the OTFS schemes~\cite{hadani2017orthogonal,hadani2018otfsnewgenerationmodulation} assuming ideal pulse-shaping waveforms that satisfy orthogonality in both time and frequency, the input-output relationship of OTFS with a practical non-orthogonal pulse shaping filter was discussed in \cite{raviteja2018practical,raviteja2018interference}.

\textcolor{black}{In~\cite{farhang2017low,surabhi2019diversity,xu2022otfs,10453468} and OFDM-based OTFS (OFDM-OTFS) architecture relying on a cyclic prefix (CP) in each transmission frame was investigated, while utilizing either a bi-orthogonal pulse~\cite{hadani2017orthogonal,surabhi2019diversity} or a rectangular shaping filter for ensuring compatibility with conventional multi-carrier systems~\cite{farhang2017low,xu2022otfs,raviteja2018interference}.} Furthermore, in~\cite{8422467}, OFDM-OTFS was incorporated into a multiple-input multiple-output (MIMO) system, and its ergodic capacity was compared to those of conventional MIMO-OFDM systems.

In addition to waveform design, channel state information (CSI) estimation was explored in OTFS.
Since the DD-domain expression of an OTFS symbol allows for a time-invariant channel representation without suffering from time-varying inter-symbol interference (ISI) in the time-domain (TD), as well as the Doppler-shift-induced inter-carrier interference (ICI) in the frequency-domain (FD),
a powerful pilot-aided joint channel estimation and data detection scheme operating in the DD-domain was developed in~\cite{raviteja2019embedded,hashimoto2021channel}.
To further improve spectral efficiency, the superposition of pilots was exploited in~\cite{yuan2021data,mishra2021otfs}.
In \cite{shen2019channel}, the three-dimensional orthogonal matching pursuit algorithm was employed for CSI estimation in a massive MIMO scenario to exploit the sparsity in the delay, Doppler, and angle domains of the channel. 
Despite the above advances, data detection remains an issue since the DD domain's support plane is wider than that of the time-frequency domain. Hence, low-complexity single-tap FD equalization of OFDM is no longer applicable. 
To get around this limitation, several message-passing (MP) detectors were proposed for OTFS~\cite{8377159,raviteja2018interference,yuan2020simple,xiang2021gaussian}, which exploit the sparsity of the DD-domain channel.
In~\cite{surabhi2019low,singh2022low,tiwari2019low,zou2021low}, the block circulant property~\cite{surabhi2019low,singh2022low} and quasi-banded sparse structure~\cite{tiwari2019low,zou2021low} of the doubly selective fading channel matrix were exploited to conceive low-complexity minimum mean squared error (MMSE)-based OTFS detectors. Furthermore, in contrast to coherent detection based on CSI estimation, a noncoherent OTFS system dispensing with pilot insertion and CSI estimation was proposed in~\cite{10453468}, which relies on blind interference cancellation in the DD domain.
\textcolor{black}{Moreover, a unified variational-inference-based receiver was developed for OTFS-based MIMO-aided integrated sensing and communication (ISAC) systems, which facilitates joint data detection, channel estimation, and kinematic parameter estimation within a probabilistic inference framework~\cite{10845803}.}

Compared to conventional time-orthogonal signaling, \textcolor{black}{the symbol interval of faster-than-Nyquist (FTN) signaling is intentionally reduced below the limit set by the Nyquist criterion.} The main benefit of FTN signaling is the compensation of the bandwidth expansion imposed by the roll-off of realistic pulse shaping in conventional time-orthogonal signaling, \textcolor{black}{thus, the transmission rate is enhanced without expanding the bandwidth requirement~\cite{ishihara2021evolution}.} Several TD~\cite{prlja2012reduced,li2017reduced}, FD~\cite{sugiura2013frequency,sugiura2014frequency} and generalized approximate MP based~\cite{ma2021generalized,ma2021parametric} demodulation schemes were proposed for mitigating the FTN-induced ISI at a low complexity.
Moreover, precoded FTN signaling schemes based on matrix factorization were proposed in~\cite{kim2016properties,ishihara2019svd,ishihara2021eigendecomposition,ishihara2022reduced}.
\textcolor{black}{Specifically, in \cite{kim2016properties}, the FTN-specific ISI channel matrix's eigenvalue decomposition (EVD) was used for transmit precoding and receiver-side weighting. In \cite{ishihara2019svd}, an EVD-based precoded FTN scheme associated with power allocation was developed for maximizing the information rate attained in additive white Gaussian noise (AWGN) channels. This was then further extended to both frequency-flat and frequency-selective fading channels in~\cite{ishihara2021eigendecomposition}. Furthermore, \cite{ishihara2022reduced} proposed a multi-carrier FTN (MFTN) scheme relying on the fast Fourier transform (FFT) for reducing the computational complexity by employing the channel matrix's circulant approximation.}
However, the above-mentioned FTN schemes assume perfect CSI, which may be challenging to obtain, especially under doubly selective fading. As a remedy, joint CSI estimation and data detection schemes were developed for open-loop FTN signaling in frequency-selective fading channels~\cite{ishihara2017iterative,shi2017frequency,wu2017hybrid,li2020joint,wen2022joint} and for a sparse code multiple access scenario~\cite{8891911}.

\textcolor{black}{Recently, several FTN schemes have been designed for doubly selective fading channels to enhance robustness against both delay spread and Doppler shift~\cite{wu2016frequency,yuan2016variational,zhou2022precoded,ishihara2023differential}.}
In~\cite{wu2016frequency}, the Gaussian MP detector, assisted by a vector form factor graph, was proposed. In~\cite{yuan2016variational}, based on variational inference, a low-complexity FD equalizer was designed for achieving a performance close to the conventional MMSE equalizer. In~\cite{zhou2022precoded}, EVD-precoded FTN signaling was designed for an acoustic doubly selective underwater channel. 
In~\cite{10902515}, an adaptive transmit precoding method for FTN signaling was proposed for fast-fading multipath channels, utilizing real-time CSI feedback from the receiver.
In~\cite{ishihara2023differential}, differential multi-carrier FTN (DMFTN) signaling was proposed, which approximately diagonalizes the FTN-specific ISI and noise correlation matrices, while dispensing with CSI estimation at the receiver.
Furthermore, pilot designs using the least sum of squared errors algorithm~\cite{keykhosravi2023pilot} and index modulation~\cite{10552123} were also proposed.
\textcolor{black}{However, the majority of prior OTFS studies assumed that the transmitted signal adheres to the time-orthogonal Nyquist criterion.}
In other words, most previous FTN studies did not consider CSI estimation and data detection in the DD domain, and the amalgamation of FTN signaling with OTFS modulation \cite{zekun2024wcl,10713220,11078288} is a hitherto neglected issue.
An exception is constituted by \cite{zekun2024wcl}, where the OTFS-based FTN signaling scheme was proposed under the assumption of Nyquist-based pilot transmission, as well as full-complexity EVD-based precoding and detection.
\textcolor{black}{An ISAC waveform was proposed by amalgamating OTFS modulation and FTN signaling while considering an ideal rectangular shaping filter, where the communication throughput and the range-velocity sensing performance were enhanced in the face of a high-mobility channel~\cite{10713220}.}
For scenarios having limited spectral resources while suffering from high Doppler shifts (ex., aeronautical communications~\cite{xu2018adaptive} and high-speed railway communication), the combination of FTN and OTFS in such contexts offers a promising solution by improving both the spectral efficiency and Doppler resilience.
\textcolor{black}{Note that the conventional OTFS scheme typically assumes having circular channel convolution in the DD domain and uncorrelated noise after matched filtering. However, these assumptions no longer hold for the OTFS-based FTN signaling scheme. As a result, classical OTFS channel estimation and detection cannot be applied.}
\begin{table*}[t]
\centering
\caption{\color{black}{Contrasting our proposed OTFS-FTN scheme with existing OTFS and FTN related works}}
\renewcommand{\arraystretch}{0.8}
\setlength{\tabcolsep}{4pt}
\begin{tabular}{|l|c|c|c|c|c|c|c|c|}
\hline
\textbf{} 
& \cite{raviteja2019embedded,hashimoto2021channel} 
& \cite{surabhi2019low,singh2022low,tiwari2019low,zou2021low} 
& \cite{ishihara2017iterative,shi2017frequency,wu2017hybrid,li2020joint,wen2022joint}
& \cite{wu2016frequency,zhou2022precoded,10902515} 
& \cite{yuan2016variational,ishihara2023differential} 
& \cite{keykhosravi2023pilot,10552123} 
& \cite{zekun2024wcl,10713220,11078288} 
& \textbf{Proposed} \\ \hline

OTFS framework & \checkmark & \checkmark &  &  &  &  &\checkmark &\checkmark \\ \hline
FTN signaling &  &  & \checkmark & \checkmark & \checkmark & \checkmark &\checkmark &\checkmark \\ \hline
Reduced-complexity detector &  & \checkmark & \checkmark  & \checkmark & \checkmark &  & &\checkmark \\ \hline
Doubly selective fading channel & \checkmark & \checkmark &  & \checkmark & \checkmark & \checkmark & \checkmark&\checkmark \\ \hline
Channel estimation and detection & \checkmark &  & \checkmark  &  &  &\checkmark  & &\checkmark \\ \hline
 \textbf{FTN-based pilot CSI estimation} &  &  &  &  &  &  & &\checkmark \\ \hline
 \textbf{Reduced-complexity detector for OTFS-FTN} &  &  &  &  &  &  & &\checkmark \\ \hline
 \textbf{DD domain input-output model for RRC filter} &  &  &  &  &  &  & &\checkmark \\ \hline
\end{tabular}
\label{tab:comparison}
\end{table*}

\textcolor{black}{
To highlight the novelty of the proposed design, Table \ref{tab:comparison} contrasts representative OTFS- and FTN-based works reported in the literature.
Against this background, the novel contributions of this paper are summarized as follows.}
\begin{itemize}
\item
We propose a DD-domain CSI estimator assisted by FTN-based pilot transmission and reduced-complexity data detection in the context of an OTFS-based FTN (OTFS-FTN) architecture conceived for doubly selective fading channels by employing a root-raised-cosine (RRC) shaping filter.\footnote{\color{black}Although RRC shaping filters are indeed routinely employed in Nyquist-rate OTFS schemes~\cite{10540354} and in orthogonal delay-Doppler division multiplexing (ODDM)~\cite{10562334} schemes, OTFS-FTN signaling has not been analytically investigated in the above literature.}
Furthermore, based on this input-output relationship and noise whitening in the DD domain, we designed bespoke FTN-based pilot (FTNP) symbols for enhancing the spectral efficiency, while attaining comparable performance to the conventional Nyquist-based pilot schemes.
\item
We propose a reduced-complexity linear MMSE (LMMSE) equalizer for the proposed OTFS-FTN system having the complexity order of $O\left[M N \log _2(N)\right]$ where sparsity approximation is employed for the FTN-induced ISI matrix.
\item \textcolor{black}{Finally, our performance analysis reveals that the proposed OTFS-FTN scheme achieves an improved BER performance and a higher information rate than all other previous FTN transmission schemes and than its Nyquist-based OTFS counterpart, while using the same RRC shaping filter.}
\end{itemize}

The remainder of this paper is organized as follows. In Section~\ref{model}, we present the system model of our proposed scheme. Section~\ref{FTNP} derives the FTNP-based channel estimator conceived while Section~\ref{LMMSE} presents the proposed LMMSE-Based OTFS-FTN receiver, followed by Section~\ref{simulation} providing our performance results.
Finally, Section~\ref{conclusion} concludes the paper.

\section{System Model}
\label{model}
\subsection{Transmit Signal}

\begin{figure*}[t]
\centering
\includegraphics[width=0.7\linewidth]{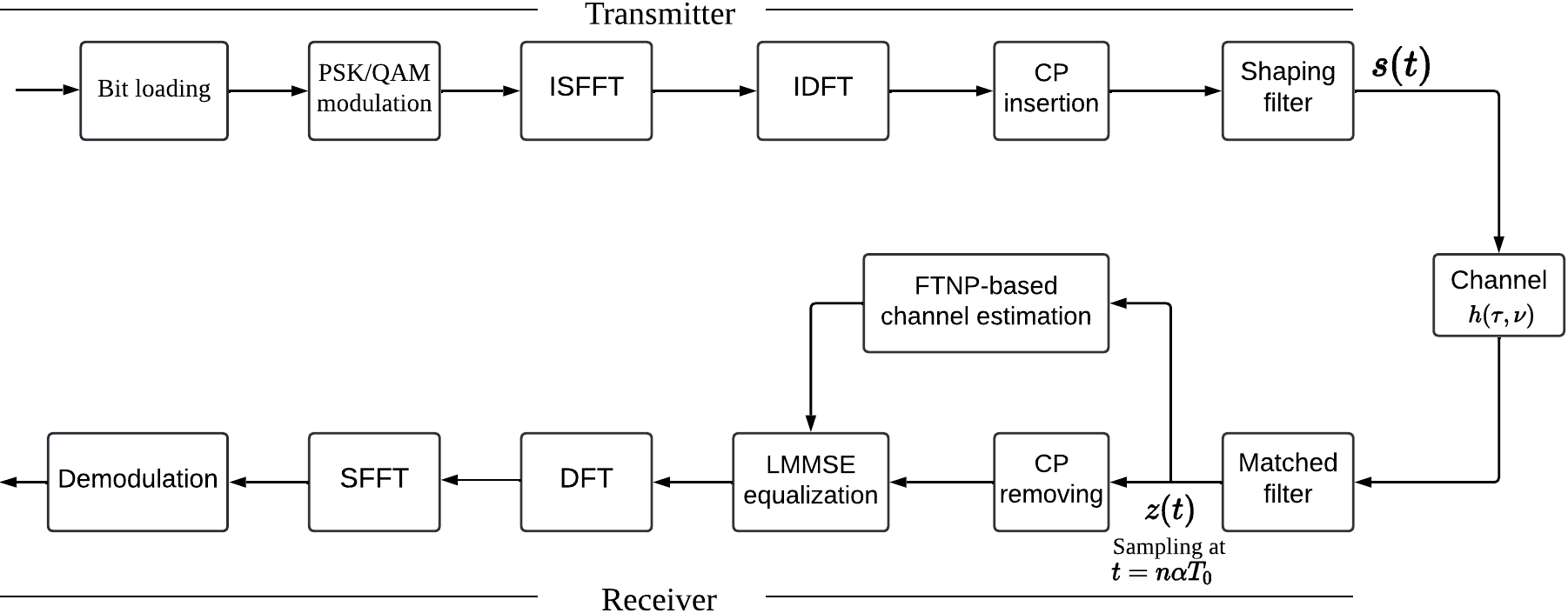}
\caption{Schematic of the proposed OTFS-FTN transceiver.}
\label{flowchart}
\end{figure*}
\textcolor{black}{Fig.~\ref{flowchart} illustrates the architecture of the proposed OTFS-FTN transceiver.
The two-dimensional DD-domain information symbols are represented by $\mathbf{X} \in {\mathbb{C}^{M \times N}}$, whose vectorized form $\mathbf{X}$ is given by $\mathbf{x}=\mathrm{vec}(\mathbf{X})=[x_{0}, \cdots, x_{MN-1}]^T \in {\mathbb{C}^{MN}}$ with an average energy of $\mathbb{E}\left[\left|x_{n}\right|^2\right]=\sigma_{x}^2 \ (n=0,\cdots, MN-1)$.}

\textcolor{black}{Upon assuming $M$ subcarriers and $N$ time slots for each transmission frame, the bandwidth and the frame interval are expressed by $M \Delta f$ and $NT$, where the subcarrier bandwidth $\Delta f$ obeys $T=1/\Delta f$.
The bandwidth of an ideal rectangular pulse filter is given by $2W$, corresponding to the Nyquist criterion-based symbol interval $T_0=1/(2W)$.
Furthermore, the FTN-specific sampling interval is given by $T_{\mathrm{f}}=\alpha T_{0}={T}/{M}$, where $\alpha$ is the packing ratio. 
\footnote{\textcolor{black}{In contrast to other OTFS-FTN schemes~\cite{11078288}, which employ time- and frequency-domain compressions and introduce coupled ICI and inter-Doppler interference (IDI), the proposed OTFS-FTN framework exploits symbol compressions only in the time domain, while preserving the orthogonality of subcarriers.}}
Similar to the conventional OFDM-OTFS scheme~\cite{xu2022otfs,10453468}, the inverse symplectic fast Fourier transform (ISFFT) is harnessed for transforming the DD-domain modulated symbols into the time-frequency domain, corresponding to the $N$-point IFFT along the rows and the $M$-point FFT along the columns in $\mathbf{X}$.
Finally, an $M$-point IFFT is used to generate the TD signal, which is denoted as}
\begin{IEEEeqnarray}{rCL}
\mathbf{S}=\mathbf{F}_M^{H}\left(\mathbf{F}_M \mathbf{X} \mathbf{F}_N^{H}\right) \in \mathbb{C}^{M \times N},\label{eq:S}
\end{IEEEeqnarray}
\textcolor{black}{where $\mathbf{F}_M \in \mathbb{C}^{M \times M}$ and $\mathbf{F}_N\in \mathbb{C}^{N \times N}$ represent the normalized discrete Fourier transform (DFT) matrices,
whose entries in the $m$th-row and $n$th-column are denoted by $\frac{1}{\sqrt{M}} e^{-2 \pi j (m-1) (n-1) / M}$ and $\frac{1}{\sqrt{N}} e^{-2 \pi j (m-1) (n-1) / N}$, respectively.}

\textcolor{black}{The column-wise vectorization of $\mathbf{S}$ is then denoted as:}
\begin{IEEEeqnarray}{rCL}
\mathbf{s}&=&\mathrm{vec}({\mathbf{S}})\in \mathbb{C}^{MN} \nonumber\\
&=&\left(\mathbf{F}_N^{H} \otimes \mathbf{I}_{M}\right) \mathbf{x},\label{eq:s}
\end{IEEEeqnarray}
where $\mathbf{I}_{M} \in \mathbb{R}^{M \times M}$ is the identity matrix, and $\otimes$ represents the Kronecker product. We assume that a $2c$-length CP is appended to the end of $\mathbf{s}$ as follows:
\begin{IEEEeqnarray}{rCL}
\mathbf{a}&=&\left[s_{0}, \ldots, s_{MN-1}, s_{0}, \ldots, s_{2 c-1}\right]^T \in {\mathbb{C}^{MN+2c}}\nonumber \\
&=&\left[a_{ 0}, \ldots, a_{MN+2 c-1}\right]^T \nonumber \\
&=&\mathbf{A}_{\mathrm{cp}} \mathbf{s},
\end{IEEEeqnarray}
where we have:
\begin{IEEEeqnarray}{rCL}
\mathbf{A}_{\mathrm{cp}}=\left[\begin{array}{ll}
\mathbf{I}_{MN} & \\
\mathbf{I}_{2 c} & \mathbf{0}_{2 c \times(MN-2 c)}
\end{array}\right] \in \mathbb{R}^{(MN+2 c) \times MN}.
\end{IEEEeqnarray}

\textcolor{black}{Furthermore, the baseband OTFS-FTN transmit signal is given by}
\begin{IEEEeqnarray}{rCL}
s(t)= \sum_{n=0}^{MN+2c-1} a_{n} h_{\mathrm{tx}}(t-nT_{\mathrm{f}}), \label{eq:st}
\end{IEEEeqnarray}
\textcolor{black}{where $h_{\mathrm{tx}}(t)$ represents an RRC pulse shaping filter with the roll-off factor $\beta$. By inserting a sufficiently long CP, the impact of FTN-induced inter-frame interference (IFI) can be neglected for practical packing ratios of $\alpha \geq1/(1+\beta)$~\cite{ishihara2022reduced,ishihara2023differential}.}

\subsection{Channel Model}
\textcolor{black}{The signal received from the time-varying channel is expressed as follows~\cite{hadani2017orthogonal}:}
\begin{IEEEeqnarray}{rCL}
r(t)=\iint h(\tau, \nu) s(t-\tau) e^{j 2 \pi \nu(t-\tau)} d \tau d \nu+w(t), \label{eq:rt}
\end{IEEEeqnarray}
where $\tau$ and $\nu$ denote the delay and Doppler shift, respectively. Furthermore, $w(t)$ represents the complex-valued AWGN component, whose power spectral density is given by $\sigma_0^2$. Furthermore, $h(\tau, \nu)$ is the channel response, which is expressed, owing to the sparsity in the DD domain, by
\begin{IEEEeqnarray}{rCL}
h(\tau, \nu)=\sum_{i=0}^{P-1}h_{i} \delta(\tau-\tau_i)\delta(\nu-\nu_i), \label{eq:h}
\end{IEEEeqnarray}
where $P$ is the number of channel taps, and $\delta(\cdot)$ is Dirac's delta function. Furthermore, $h_i$, $\tau_i$, and $\nu_i$ represent the complex-valued channel gain, the delay, and the Doppler shift of the $i$th path.
More specifically, the delay and Doppler shift in the $i$th path are denoted, respectively, by
\begin{IEEEeqnarray}{rCL}
\tau_i&=&\frac{l_i}{M \Delta f} \label{eq:delay}\\
\quad \nu_i&=&\frac{k_i+\kappa_i}{N T} \label{eq:tap},
\end{IEEEeqnarray}
where $l_i$ and $k_i$ are integer parameters corresponding to the delay and Doppler shift, and $\kappa_i$ represents the fractional Doppler shift for $-{1}/{2}<\kappa_{i} \leq {1}/{2}$. We assume that the channel's maximum delay $\tau_{\max }$ and maximum Doppler tap $\nu_{\max }$ satisfy $\tau_{\max } \leq (2c-1) T / M$ and $\left|\nu_{\max }\right| \leq \Delta f / 2$, respectively. Furthermore, $l_{\max}$ and $k_{\max}$ correspond to $\tau_{\max }$ and $\nu_{\max }$, which are calculated from \eqref{eq:delay} and \eqref{eq:tap}. 
\textcolor{black}{Note that in the DD domain, the paths having different Doppler shifts, but the same delay are separable.}

\subsection{Receive Signal}
From \eqref{eq:st}, \eqref{eq:rt}, and \eqref{eq:h}, the received signal after matched filtering using the pulse shaping filter $h_{\mathrm{rx}}(t)=h_{\mathrm{tx}}(t)$, is given by
\begin{IEEEeqnarray}{rCL}
z(t)\!&=&\!\left[\!\sum_{i=0}^{P-1} h_i e^{j 2 \pi \frac{(k_i+\kappa_i)\left(t-l_i T_\mathrm{f}\right)}{M N T_{\mathrm{f}}}} s\left(t-l_i  T_{\mathrm{f}}\right)\!+\!w(t)\!\right]\!*\! h_{\mathrm{rx}}^{*}(-t)\nonumber\\
\!&=&\!\sum_{i=0}^{P-1} \sum_{n=0}^{MN+2c-1} \!\!\!h_i e^{j 2 \pi \frac{(k_i+\kappa_i)\left(t-l_i T_{\mathrm{f} }\right)}{M N T_{\mathrm{f}}}} a_n g\left(t-\left(n+l_i\right) T_{\mathrm{f}}\right)\nonumber\\ && +\eta(t), \label{eq:rt_m}
\end{IEEEeqnarray}
where $g(t) \triangleq h_{\mathrm{tx}}(t) * h_{\mathrm{rx}}^{*}(-t)$ is the raised-cosine (RC) filter, and $\eta(t) \triangleq w(t) * h_{\mathrm{rx}}^{*}(-t)$ is the FTN-specific colored noise.

By sampling $z(t)$ at the FTN interval and removing the first $c$ and last $c$ samples, we obtain the received samples of
\begin{IEEEeqnarray}{rCL}
\mathbf{z}&=&[z_0,\cdots,z_{MN-1}]^T \ \in {\mathbb{C}^{MN}} \nonumber \\
&=&[z(c T_\mathrm{f}),\cdots,z((MN+c-1)T_\mathrm{f})]^T \nonumber \\
&=&\mathbf{R}_{\mathrm{cp}}\mathbf{H} \mathbf{A}_{\mathrm{cp}} \mathbf{s}+\boldsymbol{\eta},\label{z}
\end{IEEEeqnarray}
where we have
\begin{IEEEeqnarray}{rCL}
\mathbf{R}_{\mathrm{cp}}\!=\!\left[\begin{array}{lll}
\!\mathbf{0}_{MN \times c} \!& \!\mathbf{I}_{MN} & \mathbf{0}_{MN \times c}
\end{array}\right]\!\in \!\!\mathbb{R}^{MN \times(MN+2 c)}, \ \ \
\end{IEEEeqnarray}
and $\boldsymbol{\eta}\in {\mathbb{C}^{MN}}$ is the FTN-specific correlated AWGN, having a correlation matrix $\mathbb{E}\left[\boldsymbol{\eta}\boldsymbol{\eta}^H\right]=\sigma_0^2 \mathbf{G}$~\cite{ishihara2019svd,ishihara2021eigendecomposition}.
Moreover, $\mathbf{G} \in \mathbb{R}^{MN \times MN}$ is a Toeplitz matrix, \textcolor{black}{the first row of which is given by $[g(0), g(-T_{\mathrm{f}}) \cdots, g(-(MN-1)T_{\mathrm{f}})]$.
Furthermore, $\mathbf{H} \in \mathbb{C}^{(MN+2c) \times (MN+2c)}$ represents the effective channel matrix, accounting for both the FTN-induced ISI and for the dispersive channel effects.
More specifically, the entry in the $k$th-row and $m$th-column of $\mathbf{H}$ is given as follows:}
\begin{IEEEeqnarray}{rCL}
{H}[k,m]\!\!=\!\!\sum_{i=0}^{P-1} \!h_i e^{j 2 \pi \frac{(k_i+\kappa_i)\left(k-l_i\right)}{M N}} \!g\!\left(k T_{\mathrm{f}}\!-\!\left(m+l_i\right) T_{\mathrm{f}}\right). 
\end{IEEEeqnarray}

\section{FTNP-Based Channel Estimation}
\label{FTNP}
\begin{figure}[t]
\centering
\includegraphics[width=0.6\linewidth]{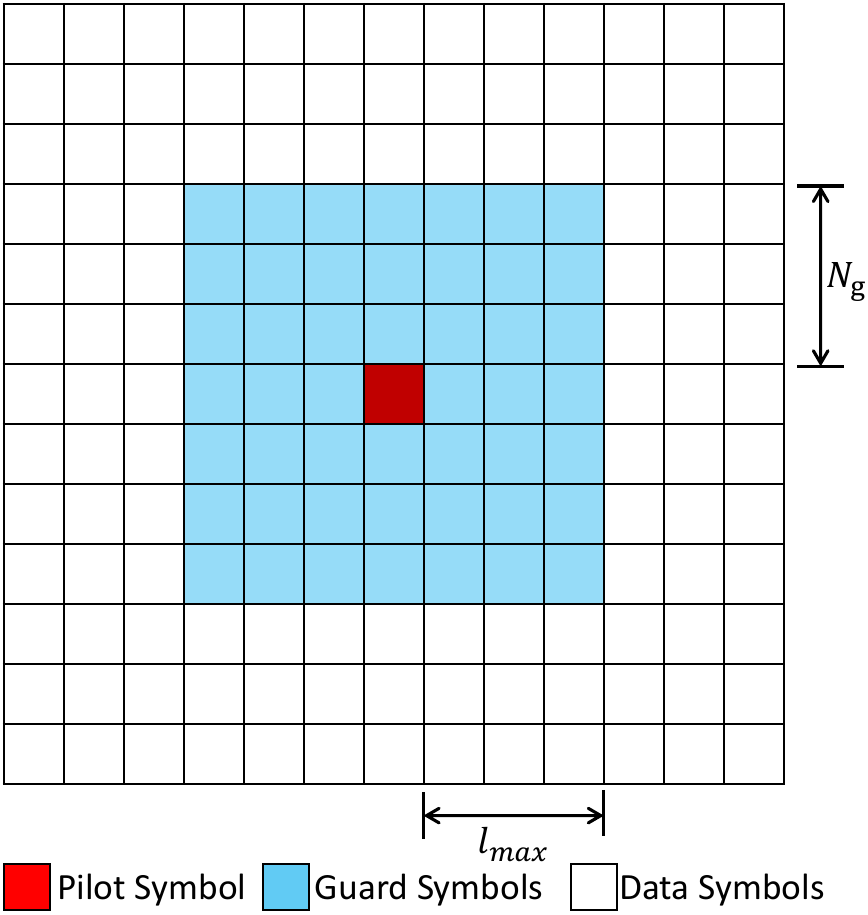}
\caption{DD-domain frame structure of $\mathbf{X}$, including the pilot symbol, guard symbols, and data symbols.}
\label{pilot}
\end{figure}
As shown in Fig.~\ref{pilot}, the symbols of each frame $\widetilde{\mathbf{X}}=\mathbf{X}^T \in {\mathbb{C}^{N \times M}}$ transmitted in the DD-domain are divided into three groups, namely the pilot symbol $\widetilde{x}_p$, the zero-padded guard symbols, and the data symbols. 
\textcolor{black}{The entry in the $k$th-row and $l$th-column of $\widetilde{\mathbf{X}}$ is denoted as \footnote{\color{black}In contrast to the conventional Nyquist-based pilot-aided OTFS channel estimation~\cite{raviteja2019embedded}, the proposed scheme is developed for OTFS-based FTN signaling relying on an RRC shaping filter, hence introducing additional coupling and correlations in the DD-domain input-output relationship. Consequently, the conventional pilot signal model and the associated channel estimation of~\cite{raviteja2019embedded} cannot be directly applied to the proposed OTFS-FTN scheme.}}
\begin{IEEEeqnarray}{rCL}
\label{eq:h_e}
\widetilde{X}[k,l] = \begin{cases}\widetilde{x}_\mathrm{p} & k=k_0, l=l_0 \\ 0 & k_0-N_\mathrm{g} \leq k \leq k_0+N_\mathrm{g} \\ & l_0-l_{\max} \leq l \leq l_0+l_{\max} \\ \textrm{data~symbols} & \text{otherwise }\end{cases}, \ \ \
\end{IEEEeqnarray}
where $N_\mathrm{g}$ is the size of guard symbols, which satisfies $N_\mathrm{g} \geq 2k_{\max}$.

\textit{Theorem 1}: Let us denote the two-dimensional symbols received in the DD domain by $\widetilde{\mathbf{Y}} \in {\mathbb{C}^{N \times M}}$ corresponding to $\widetilde{\mathbf{X}}$.
By ignoring the effects of AWGNs, \textcolor{black}{the entry in the $k$th-row and $l$th-column of $\widetilde{\mathbf{Y}}$ is expressed as}
\begin{IEEEeqnarray}{rCL}
\widetilde{Y}[k, l]& \approx &\sum_{i=0}^{P-1} \sum_{q=-N_i}^{N_i} h_i e^{j 2 \pi\frac{(l-l_{i})(k_{i}+\kappa_{i})}{MN}} \gamma_i(k, l, q) \nonumber \\
&& \times \widetilde{X}\left[\left[k-k_{i}+q\right]_N,\left[l-l_{i}\right]_M\right], \label{DD1}
\end{IEEEeqnarray}
where we have:
\begin{IEEEeqnarray}{rCL}
&&\gamma_i(k, l, q)=\nonumber \\ &&\begin{cases}\frac{T}{N}\!g((l-l_i)T_{\mathrm{f}})\rho(q,\kappa_i) \!& \!l_{i} \leq \!l\!<\!M \\ \frac{T}{N}\!g((l-l_i+M)T_{\mathrm{f}})(\rho(q,\kappa_i)\!-\!1) e^{-j 2 \pi \frac{\left[k-k_{i}+q\right]_N}{N}} \!\!\!\!& \!0 \leq\! l\!<l_{i}\end{cases}\nonumber \\ \label{DD2}
\end{IEEEeqnarray}
\begin{IEEEeqnarray}{rCL}
\rho(q,\kappa_i)=\frac{e^{-j 2 \pi\left(-q-\kappa_{i}\right)}-1}{e^{-j \frac{2 \pi}{N}\left(-q-\kappa_{i}\right)}-1}. \label{DD3}
\end{IEEEeqnarray}
\textit{Proof}: The proof is given in Appendix A.
{\color{black}According to Theorem~1, the non-orthogonality of the RRC pulses under FTN signaling introduces off-diagonal coupling terms in the DD domain, which results in a non-circulant, but structured channel matrix.}

Let us assume that $\hat{P}$, $\hat{l}_i$, $\hat{k}_i$, $\hat{\kappa}_i$, $\hat{h}_i$ are the estimated values of the total number of channel taps, delay, integer Doppler tap, fractional Doppler tap, and channel coefficient, respectively.
Then, our FTNP-based channel estimator is constituted by the following four steps:\\
\textit{1) Delay estimation:}
\textcolor{black}{After the $M$-point FFT and SFFT operation, the samples in \eqref{z} are converted into those of the DD-domain, which are given by}
\begin{IEEEeqnarray}{rCL}
\mathbf{y}&=&\left(\mathbf{F}_N \otimes \mathbf{I}_{M}\right) \mathbf{z}\in \mathbb{C}^{MN} \nonumber \\
&=&\left(\mathbf{F}_N \otimes \mathbf{I}_{M}\right) \mathbf{R}_{\mathrm{cp}}\mathbf{H} \mathbf{A}_{\mathrm{cp}} \mathbf{s}+\boldsymbol{\eta}_\mathrm{d},
\end{IEEEeqnarray}
where we have $\boldsymbol{\eta}_\mathrm{d}=\left(\mathbf{F}_N \otimes \mathbf{I}_{M}\right)\boldsymbol{\eta}\in \mathbb{C}^{MN}$, denoting the DD-domain's equivalent correlated noise and the corresponding covariance matrix is formulated as
\begin{IEEEeqnarray}{rCL}
\mathbb{E}\left[\boldsymbol{\eta}_\mathrm{d} \boldsymbol{\eta}_\mathrm{d}^H\right]&=&\left(\mathbf{F}_N \otimes \mathbf{I}_M\right)\mathbb{E}[{\boldsymbol{\eta \eta}^H}] \left(\mathbf{F}_N^H \otimes \mathbf{I}_M\right)\nonumber \\
&=&\sigma_0^2\left(\mathbf{F}_N \otimes \mathbf{I}_M\right)\mathbf{G}\left(\mathbf{F}_N^H \otimes \mathbf{I}_M\right)\nonumber \\
&=&\mathbf{G}_\mathrm{d},
\end{IEEEeqnarray}
where $\mathbf{G}_\mathrm{d}=\sigma_0^2 \left(\mathbf{F}_N \otimes \mathbf{I}_M\right)\mathbf{G}\left(\mathbf{F}_N^H \otimes \mathbf{I}_M\right) \in \mathbb{C}^{MN \times MN}$.
We then harness the noise whitening to reduce the effects of DD-domain correlated noise, which is carried out by multiplying the weight matrix as follows:
\begin{IEEEeqnarray}{rCL}
\mathbf{y}_\mathrm{w}=\mathbf{G}_\mathrm{d}^{-\frac{1}{2}} \mathbf{y}\in \mathbb{C}^{MN},
\end{IEEEeqnarray}
where $\mathbf{y}_\mathrm{w}$ is reshaped to $\widetilde{\mathbf{Y}}_\mathrm{w}\in {\mathbb{C}^{N \times M}}$ by exploiting the relationship of $\mathbf{y}_\mathrm{w}=\mathrm{vec}(\widetilde{\mathbf{Y}}_\mathrm{w})$.
To estimate the number of channel taps, we introduce a threshold $\mathcal{T}$, which is based on hypothesis testing~\cite{van2004detection}
\begin{IEEEeqnarray}{rCL}
\label{T}\mathcal{T}=1+\sqrt2Q^{-1}(P_{fa}),
\end{IEEEeqnarray}
where $Q^{-1}(P_{fa})$ is the inverse of the Gaussian Q-function and $P_{fa}$ is the false alarm rate.
Let us define the $k$th-row and $l$th-column entry of $\widetilde{\mathbf{Y}}_\mathrm{w}$ as $\widetilde{Y}_\mathrm{w}[k,l]$. Then,
if $\widetilde{Y}_\mathrm{w}[k,l]$ satisfies $|\widetilde{Y}_\mathrm{w}[k,l]|^2>\mathcal{T}$ within the guard interval defined by $k_0-N_\mathrm{g} \leq k \leq k_0+N_\mathrm{g}$, $l_0\leq l \leq l_0+l_{\max}$, a propagation path corresponding to $\hat{l}_i=l-l_0$ exists for each $(k,l)$ pair.
The number of channel taps $\hat{P}$ is estimated by the total number of paths given above.

\textit{2) Doppler shift estimation:}
Similar to the rectangular pulse scenario of~\cite{raviteja2018interference}, observe in \eqref{DD2} that $\gamma_i(k, l, q)$ reaches its maximum value at $q=0$.
Hence, for each estimated path $\widetilde{Y}[k_0+k, l_0+\hat{l}_i]$ ($i=0,\cdots,\hat{P}-1$), the integer Doppler tap is estimated by
\begin{IEEEeqnarray}{rCL}
\hat{k}_i= \underset {\forall k \in[-k_\textrm{max}, k_\textrm{max}]}{\arg \max}\left|\widetilde{Y}[k_0+k,l_0+\hat{l}_i]\right|^2\!\!.
\end{IEEEeqnarray}

\textit{3) Fractional Doppler shift estimation:}
Based on \eqref{DD1}, \eqref{DD2}, and \eqref{DD3}, $\widetilde{Y}[k_0+\hat{k}_i-1,l_0+\hat{l}_i]$ and $\widetilde{Y}[k_0+\hat{k}_i,l_0+\hat{l}_i]$ are given by
\begin{IEEEeqnarray}{rCL}
&&\widetilde{Y}[k_0+\hat{k}_i-1,l_0+\hat{l}_i] \approx  h_i e^{j 2 \pi\frac{l_0(k_{i}+\kappa_{i})}{MN}} \frac{T}{N}g(l_0 T_\mathrm{f}) \nonumber \\
&& \times \rho(1,\kappa_i) \widetilde{X}\left[k_0,l_0\right] \\
&&\widetilde{Y}[k_0+\hat{k}_i,l_0+\hat{l}_i] \approx  h_i e^{j 2 \pi\frac{l_0(k_{i}+\kappa_{i})}{MN}} \frac{T}{N}g(l_0 T_\mathrm{f})\nonumber \\
&& \times \rho(0,\kappa_i) \widetilde{X}\left[k_0,l_0\right],
\end{IEEEeqnarray}
where the only difference is represented by the fading components $\rho(1,\kappa_i)$ and $\rho(0,\kappa_i)$.
Therefore, for the $i$th channel tap, the fractional Doppler index is estimated for the predefined precision of $\kappa_i$ as follows:
\begin{IEEEeqnarray}{rCL}
\hat{\kappa}_i\!\!=\!\!\underset{\forall \kappa \in\left(-\frac{1}{2}, \frac{1}{2}\right]}{\arg \min }\!\left|\frac{\widetilde{Y}[k_0+\hat{k}_i,l_0+\hat{l}_i]}{\widetilde{Y}[k_0+\hat{k}_i-1,l_0+\hat{l}_i]}\!-\!\frac{\rho(0,\kappa_i)}{\rho(1,\kappa_i)}\right|^2\!\!.
\end{IEEEeqnarray}

\textit{4) Channel coefficient estimation:}
Based on \eqref{DD1}--\eqref{DD3}, the fading coefficient corresponding to the $i$th tap is estimated by
\begin{IEEEeqnarray}{rCL}
\hat{h}_i=\frac{\widetilde{Y}[k_0+\hat{k}_i,l_0+\hat{l}_i]}{E_{\mathrm{p}}\frac{T}{N}e^{j 2 \pi \frac{l_0\left(\hat{k}_i+\hat{\kappa}_i\right)}{M N}} g(l_0 T_\mathrm{f})\rho(0,\hat{\kappa}_i)},
\end{IEEEeqnarray}
where $E_{\mathrm{p}}$ is the power of a pilot symbol.

\section{LMMSE-Based OTFS-FTN Receiver}
\label{LMMSE}

\subsection{Low-Complexity LMMSE Equalizer}
\begin{figure}[t]
\centering
\includegraphics[width=\linewidth]{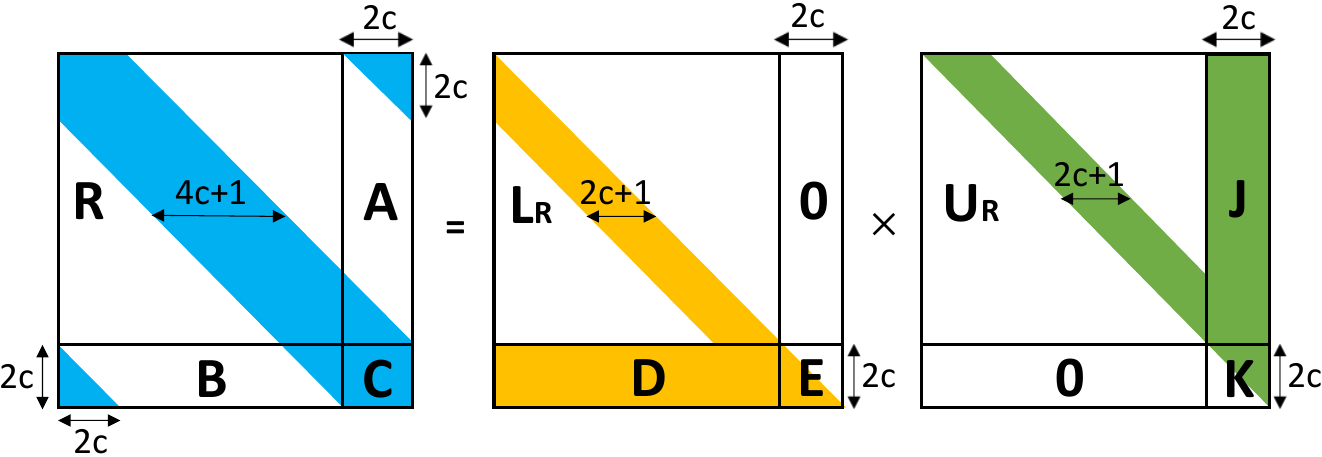}
\caption{The structure of LU decompositon for $\mathbf{W}_1$.}
\label{lu}
\end{figure}
Firstly, let us consider LMMSE equalization along with noise whitening for the received signals of our OTFS-FTN scheme in~\eqref{z} as follows:~\cite{tiwari2019low,surabhi2019low,zou2021low}
\begin{IEEEeqnarray}{rCL}
\hat{\mathbf{s}}_\textrm{MMSE}&=&(\mathbf{H_\mathrm{t} Q})^{H}\left[(\mathbf{H_\mathrm{t} Q})(\mathbf{H_\mathrm{t} Q})^{H}+{\sigma_0^2} \mathbf{G}\right]^{-1} \mathbf{z} \nonumber \\
&=&\mathbf{Q}^{H} \mathbf{H}_\mathrm{t}^{H}\left[\mathbf{H}_\mathrm{t} \mathbf{H}_\mathrm{t}^{H}+{\sigma_0^2} \mathbf{G}\right]^{-1} \mathbf{z},\label{zlmmse_act}
\end{IEEEeqnarray}
where $\mathbf{Q}=\mathbf{F}_N^{H} \otimes \mathbf{I}_{M}$ and $\mathbf{H}_\mathrm{t}=\mathbf{R}_{\mathrm{cp}}\mathbf{H} \mathbf{A}_{\mathrm{cp}}$.
Note that the calculations of \eqref{zlmmse_act} have a complexity order as high as $O\left[M^3 N^3\right]$.

\textcolor{black}{ISI in the proposed OTFS-FTN scheme relying on the RRC shaping filtering exhibits a structured and localized DD-domain channel matrix. Specifically, interference is confined to a limited range, as determined by the FTN packing ratio and the roll-off factor of the RRC shaping filter. 
This property results in sparsity in the effective channel matrix, which can be exploited for reduced-complexity detection.}
To reduce the complexity, under the conditions of $g(nT_{\mathrm{f}}) \approx 0$ for $|n|>c$ with a sufficiently long CP,
the received frame $\eqref{z}$ is approximated by
\begin{IEEEeqnarray}{rCL}
\label{aprHs}\mathbf{z}\approx\mathbf{H}_\mathrm{s} \mathbf{s}+\boldsymbol{\eta},
\end{IEEEeqnarray}
where we consider a sparse matrix $\mathbf{H}_\mathrm{s} \in \mathbb{C}^{MN\times MN}$ to approximate $\mathbf{R}_{\mathrm{cp}}\mathbf{H} \mathbf{A}_{\mathrm{cp}}$.
More specifically, for $k\leq MN-2c$, the $k$th-row of $\mathbf{H}_\mathrm{s}$ is given by
\begin{IEEEeqnarray}{rCL}
&[\underbrace{0, \ldots, 0}_{k-1},
\sum_{i=0}^{P-1} h_i e^{j 2 \pi \frac{\left({k}_i+{\kappa}_i\right)(k+c-l_i)}{M N}}g((-c+l_i) T_\mathrm{f}), \ldots,\nonumber \\
& \sum_{i=0}^{P-1} h_i e^{j 2 \pi \frac{\left({k}_i+{\kappa}_i\right)(k+c-l_i)}{M N}}g((c+l_i) T_\mathrm{f}), \underbrace{0, \ldots, 0}_{MN-2 c-k}],
\end{IEEEeqnarray}
and for $k>MN-2 c$, the $k$th-row of $\mathbf{H}_{\mathrm{s}}$ is formulated as
\begin{IEEEeqnarray}{rCL}
&[\sum_{i=0}^{P-1} h_i e^{j 2 \pi \frac{\left({k}_i+{\kappa}_i\right)(k+c-l_i)}{M N}}g((MN-c-k+l_i+1) T_\mathrm{f}), \ldots, \nonumber \\
&\sum_{i=0}^{P-1} h_i e^{j 2 \pi \frac{\left({k}_i+{\kappa}_i\right)(k+c-l_i)}{M N}} g((c+l_i) T_\mathrm{f}),
\underbrace{0, \ldots, 0}_{MN-2 c-1},\nonumber \\
&\sum_{i=0}^{P-1} h_i e^{j 2 \pi \frac{\left({k}_i+{\kappa}_i\right)(k+c-l_i)}{M N}}g((-c+l_i) T_\mathrm{f}), \ldots, \nonumber \\
&\sum_{i=0}^{P-1} h_i e^{j 2 \pi \frac{\left({k}_i+{\kappa}_i\right)(k+c-l_i)}{M N}}g((MN-c-k+l_i) T_\mathrm{f})].
\end{IEEEeqnarray}

By employing the above-mentioned LMMSE equalizer combined with noise whitening for the approximated received frame, we obtain the demodulated symbols as follows:
\begin{IEEEeqnarray}{rCL}
\hat{\mathbf{s}}&=&(\mathbf{H_\mathrm{s} Q})^{H}\left[(\mathbf{H_\mathrm{s} Q})(\mathbf{H_\mathrm{s} Q})^{H}+{\sigma_0^2} \mathbf{G}\right]^{-1} \mathbf{z} \nonumber \\
&=&\mathbf{Q}^{H} \mathbf{H}_\mathrm{s}^{H}\left[\mathbf{H}_\mathrm{s} \mathbf{H}_\mathrm{s}^{H}+{\sigma_0^2} \mathbf{G}\right]^{-1} \mathbf{z},\label{zlmmse2}
\end{IEEEeqnarray}
where by assuming sufficiently large $MN$, the Toeplitz matrix $\mathbf{G}$ can be approximated by a circulant matrix $\mathbf{G}_\mathrm{c}$~\cite{gray2006toeplitz}. 
\textcolor{black}{More specifically, the first column of $\mathbf{G}_\mathrm{c}$ can be denoted by $[g(0), g(T_\mathrm{f}), \cdots, g(2c T_\mathrm{f}), 0, \cdots, 0, g(-2c T_\mathrm{f}), \cdots, g(-T_\mathrm{f})]^T$.}
Then, \eqref{zlmmse2} is approximated by
\begin{IEEEeqnarray}{rCL}
\hat{\mathbf{s}}\approx\mathbf{Q}^{H} \mathbf{H}_\mathrm{s}^{H}\mathbf{W}_1^{-1} \mathbf{z},\label{zlmmse}
\end{IEEEeqnarray}
where we have:
\begin{IEEEeqnarray}{rCL}
\label{W_1}\mathbf{W}_1=\mathbf{H_\mathrm{s} H_\mathrm{s}}^{H}+{\sigma_0^2} \mathbf{G}_\mathrm{c}.
\end{IEEEeqnarray}
As shown in Fig. \ref{lu}, we consider the lower-upper (LU) decomposition\footnote{\color{black}{While the LU decomposition and the related forward substitution technique are also employed in conventional OTFS detectors~\cite{tiwari2019low}, the detection problem addressed here is different since the introduction of FTN signaling destroys the circular convolution structure exploited in conventional OTFS detection.
More specifically, in the proposed OTFS-FTN scheme, the effective channel matrix incorporates FTN-induced ISI and correlated additive noise.}} of $\mathbf{W}_1$ of \cite{golub2013matrix}
\begin{IEEEeqnarray}{rCL}
\underbrace{\left[\begin{array}{cc}
\mathbf{R} \!& \!\mathbf{A} \\
\mathbf{B} \!& \!\mathbf{C}
\end{array}\right]}_{\mathbf{W}_1}\!=\!\underbrace{\left[\begin{array}{ll}
\mathbf{L}_{\mathrm{R}}\! & \!\mathbf{0}_{(MN-2c) \times 2c} \\
\mathbf{D} \!& \!\mathbf{E}
\end{array}\right]}_{\mathbf{L}} \!\times\! \underbrace{\left[\begin{array}{cc}
\mathbf{U}_{\mathrm{R}} \!&\! \mathbf{J} \\
\mathbf{0}_{2c \times (MN-2c)} \!& \!\mathbf{K}
\end{array}\right]}_{\mathbf{U}},\nonumber \\
\end{IEEEeqnarray}
where we have:
\begin{IEEEeqnarray}{rCL}
\mathbf{R} & =&\mathbf{L}_\mathrm{R} \mathbf{U}_\mathrm{R} \in \mathbb{C}^{(MN-2c) \times(MN-2 c)} \label{lrur}\\
\mathbf{J} & =&\mathbf{L}_\mathrm{R}^{-1} \mathbf{A} \in \mathbb{C}^{(MN-2c) \times 2 c} \label{lr}\\
\mathbf{D} & =&\mathbf{B} \mathbf{U}_\mathrm{R}^{-1} \in \mathbb{C}^{2c \times (MN-2 c)} \label{ur}\\
\mathbf{E K} & =&\mathbf{C}-\mathbf{D J} \in \mathbb{C}^{2c \times 2c},
\end{IEEEeqnarray}
and
$\mathbf{L}_{\mathrm{R}}, \mathbf{U}_{\mathrm{R}} \in \mathbb{C}^{(MN-2c) \times(MN-2 c)}$,
$\mathbf{A} \in \mathbb{C}^{(MN-2c) \times 2 c}$,
$\mathbf{B} \in \mathbb{C}^{2c \times (MN-2 c)}$, and
$\mathbf{E}, \mathbf{K}\in \mathbb{C}^{2c \times 2c}$.
\textcolor{black}{Moreover, for the banded matrix $\mathbf{R}$, its LU decomposition can be computed efficiently by utilizing the low-complexity algorithm of~\cite{golub2013matrix}.}
Similar to \cite{tiwari2019low}, $\mathbf{J} =\mathbf{L}_\mathrm{R}^{-1} \mathbf{A}$ is calculated through the forward substitution algorithm for the lower triangular banded matrix, as shown in Algorithm~\ref{alg1}.
Furthermore, since we have the Hermitian transpose of $\mathbf{D}$ as $\mathbf{D}^H=(\mathbf{B} \mathbf{U}_\mathrm{R}^{-1})^H=(\mathbf{U}_\mathrm{R}^H)^{-1}\mathbf{B}^H$, $\mathbf{D}^H$ can be computed in a similar manner to Algorithm~\ref{alg1}.
Since $2c \ll M N$, the LU decomposition of $\mathbf{E}$ and $\mathbf{K}$ doesn't substantially increase the total complexity.
\begin{algorithm}[t]
\caption{Computation of $\mathbf{J}=\mathbf{L}_\mathrm{R}^{-1} \mathbf{A}$}
\label{alg1}
\textbf{Input}:  $\mathbf{L}_\mathrm{R}$ and $\mathbf{A}$\\
\textbf{Output}: $\mathbf{J}$\\
\textbf{for} $j =1 : 2c$ \\
\hspace{1em}$\mathbf{J}(1,j)=\mathbf{A}(1,j)/\mathbf{L}_\mathrm{R}(1,1)$\\
\textbf{end} \\
\textbf{for} $i =2 : (MN-2c)$ \\
\hspace{1em}\textbf{for} $j =1 : 2c$ \\
\hspace{2em}$\mathbf{J}(i,j)=\frac{\left(\mathbf{A}(i,j)-\sum_{m=\max (1, i-2c)}^{i-1} \mathbf{L}_\mathrm{R}(i,m) \mathbf{J}(m,j)\right)}{\mathbf{L}_\mathrm{R}(i,i)}$\\
\hspace{1em}\textbf{end}\\
\textbf{end}
\end{algorithm}

Hence, through the LU decomposition, $\hat{\mathbf{s}}$ is rewritten by
\begin{IEEEeqnarray}{rCL}
\hat{\mathbf{s}}=\mathbf{Q}^{H} \underbrace{\mathbf{H}_\mathrm{s}^{H}  \overbrace{\mathbf{U}^{-1}\underbrace{ \mathbf{L}^{-1} \mathbf{z}}_{\mathbf{W}_2}} ^{\mathbf{W}_3}}_{\mathbf{W}_4},
\end{IEEEeqnarray}
where $\mathbf{W}_2=\mathbf{L}^{-1} \mathbf{z} \in \mathbb{C}^{MN}$ is computed by the forward substitution procedure of Algorithm \ref{alg2}. Furthermore, $\mathbf{W}_3=\mathbf{U}^{-1} \mathbf{W}_2\in \mathbb{C}^{MN}$ is calculated by the low-complexity backward substitution procedure of Algorithm \ref{alg33}. Owing to the sparsity of $\mathbf{H}_\mathrm{s}^{H}$, the complexity order of $\mathbf{W}_4=\mathbf{H}_\mathrm{s}^{H} \mathbf{W}_3$ is as low as $O\left[M Nc\right]$.
Finally, the computation of $\mathbf{Q}^{H} \mathbf{W}_4$ corresponds to the $M$ FFT operation with the size of $N$, and hence the associated complexity order becomes $O\left[M N\log_2(N)\right]$.
{\textcolor{black}{While the conventional MIMO-OTFS-FTN system of~\cite{11083537} relied on similar reduced-complexity LU-decomposition-assisted detection, this paper focuses our attention on a single stream OTFS-FTN framework. 
Owing to the absence of spatial coupling in the proposed scheme, the effective channel matrix can be LU-decomposed without any elaborate matrix reordering and permutation, unlike~\cite{11083537}, which reduces the computational complexity.}}

In summary, the computational complexity of each major computational step for the proposed LMMSE-based OTFS-FTN receiver is listed in Table \ref{tab1}. Hence, the total complexity of the proposed receiver is $O[M N\left(\log_2(N)+c^2\right)]$. 
{\color{black}Table~\ref{tab:complexity} compares the computational complexity of the representative OTFS and OTFS-FTN detectors. The EVD-based detector~\cite{zekun2024wcl,11078288} exhibits a prohibitive cubic complexity of $\mathcal{O}(M^3N^3)$ due to large-scale eigenvalue decomposition.
The conventional LMMSE-based detector reduces the complexity to $\mathcal{O}(MN\log_2 N)$ by exploiting the block-circulant structure of the OTFS channel~\cite{surabhi2019low,tiwari2019low}.
The MP detector~\cite{raviteja2018interference} achieves linear scaling with respect to the frame size and the number of delay taps, but incurs an additional dependence on the modulation order $Q_b$ and the number of iterations $N_I$, resulting in a complexity of $\mathcal{O}(N_I MN P^2 Q_b)$.
Similarly, the maximal ratio combining (MRC) detector~\cite{10173746} exhibits a complexity of $\mathcal{O}\left(N_I M_{\mathrm{eff}}\, N\, G_o\, P\right)$, which scales linearly with the oversampling factor $G_o$ and the effective delay dimension $M_{\mathrm{eff}}$.
By contrast, the proposed detector exploits the sparse and banded structure of the OTFS-FTN channel matrix, enabling LU-based equalization at a reduced complexity of $\mathcal{O}\!\left[MN(\log_2 N + c^2)\right]$.
Notably, the proposed approach is non-iterative and does not depend on the modulation order.}
\begin{algorithm}[t]
\caption{Computation of $\mathbf{W}_2=\mathbf{L}^{-1} \mathbf{z}$}
\label{alg2}
\textbf{Input}:  $\mathbf{L}$ and $\mathbf{z}$\\
\textbf{Output}: $\mathbf{W}_2$\\
$\mathbf{W}_2(1)=\mathbf{z}(1)$\\
\textbf{for} $n =2 : (2c+1)$ \\
\hspace{1em}$\mathbf{W}_2(n)=\mathbf{z}(n)-\sum_{i=1}^{n-1} \mathbf{z}(n-i)\mathbf{L}(n,n-i)$\\
\textbf{end} \\
\textbf{for} $n =(2c+2) : (MN-2c)$ \\
\hspace{1em}$\mathbf{W}_2(n)=\mathbf{z}(n)-\sum_{i=1}^{2c} \mathbf{z}(n-i)\mathbf{L}(n,n-i)$\\
\textbf{end} \\
\textbf{for} $n =(MN-2c+1) : MN$ \\
\hspace{1em}$\mathbf{W}_2(n)=\mathbf{z}(n)-\sum_{i=1}^{n-1} \mathbf{z}(n-i)\mathbf{L}(n,n-i)$\\
\textbf{end}
\end{algorithm}
\begin{algorithm}[t]
\caption{Computation of $\mathbf{W}_3=\mathbf{U}^{-1} \mathbf{W}_2$}
\label{alg33}
\textbf{Input}:  $\mathbf{U}$ and $\mathbf{W}_2$\\
\textbf{Output}: $\mathbf{W}_3$\\
$\mathbf{W}_3(MN)=\mathbf{W}_2(MN)/\mathbf{U}(MN,MN)$\\
\textbf{for} $n =(MN-1) :-1: (MN-4c)$ \\
\hspace{1em}$\mathbf{W}_3(n)=(\mathbf{W}_2(n)-\sum_{i=1}^{MN-n} \mathbf{W}_3(n+i)\mathbf{U}(n,n+i))/\mathbf{U}(n,n)$\\
\textbf{end} \\
\textbf{for} $n =(MN-4c-1) : -1:1$ \\
\hspace{1em}$\mathbf{W}_3(n)=(\mathbf{W}_2(n)-\sum_{i=1}^{2c} \mathbf{W}_3(n+i)\mathbf{U}(n,n+i)
-\sum_{i=MN-2c+1}^{MN} \mathbf{W}_3(i)\mathbf{U}(n,i))/\mathbf{U}(n,n)$\\
\textbf{end}
\end{algorithm}
\begin{table}
\footnotesize
\centering
\caption{Complexity Analysis for LMMSE-Based OTFS-FTN Receiver}
\begin{tabular}{|c|c|}
\hline
\textbf{Operation} & \textbf{Computational Complexity} \\ \hline
\eqref{lrur} & $O\left(M Nc\right)$  \\ \hline
\eqref{lr} and \eqref{ur} &$O\left(M Nc^2\right)$  \\ \hline
$\mathbf{L}^{-1} \mathbf{z}$ &$O\left(M Nc\right)$  \\ \hline
$\mathbf{U}^{-1} \mathbf{W}_2$ &$O\left(M Nc\right)$  \\ \hline
$\mathbf{H}_\mathrm{s}^{H} \mathbf{W}_3$ &$O\left(M Nc\right)$  \\ \hline
$\mathbf{Q}^{H} \mathbf{W}_4$ &$O\left(M N\log_2(N)\right)$\\ \hline
\end{tabular}
\label{tab1}
\end{table}
\begin{table}[t]
\centering
\caption{Computational complexity comparison of representative OTFS/OTFS-FTN detection methods.}
\label{tab:complexity}
\renewcommand{\arraystretch}{1.1}
\setlength{\tabcolsep}{6pt}
\begin{tabular}{|l|c|}
\hline
\textbf{Detection method} & \textbf{Computational complexity} \\
\hline
\textbf{Proposed } &
$O[M N\left(\log_2(N)+c^2\right)]$ \\
\hline
LMMSE-based detector~\cite{surabhi2019low,tiwari2019low} &
$\mathcal{O}(MN\log_2 N)$ \\
\hline
MP detector ~\cite{raviteja2018interference} &
$O(N_I MN P^2 Q_b)$ \\
\hline
MRC detector~\cite{10173746} &
$\mathcal{O}\left(N_I M_{\mathrm{eff}}\, N\, G_o\, P\right)$ \\
\hline
EVD-based detector~\cite{zekun2024wcl,11078288} &
$O\left(M^3 N^3\right)$  \\
\hline
\end{tabular}
\end{table}

\subsection{Achievable Information Rate}
Let us define the weight matrix in \eqref{zlmmse} as $\mathbf{W}=\mathbf{Q}^{H} \mathbf{H}_\mathrm{s}^{H}\left[\mathbf{H}_\mathrm{s} \mathbf{H}_\mathrm{s}^{H}+{\sigma_0^2} \mathbf{G}_\mathrm{c}\right]^{-1}$. Then, \eqref{zlmmse} is rewritten as:
\begin{IEEEeqnarray}{rCL}
\hat{\mathbf{s}}&=&\mathbf{W}\mathbf{R}_{\mathrm{cp}}\mathbf{H} \mathbf{A}_{\mathrm{cp}}\mathbf{s}+\mathbf{W}\boldsymbol{\eta}\nonumber \\
&=&\mathbf{W}\mathbf{R}_{\mathrm{cp}}\mathbf{H} \mathbf{A}_{\mathrm{cp}} \mathbf{Q}\mathbf{x}+\mathbf{W}\boldsymbol{\eta}\nonumber\\
&=&\mathbf{H}_\mathrm{eff} \mathbf{x}+\boldsymbol{\eta}_\mathrm{eff},
\end{IEEEeqnarray}
where we have:
\begin{IEEEeqnarray}{rCL}
\mathbf{H}_\mathrm{eff}&=&\mathbf{W}\mathbf{R}_{\mathrm{cp}}\mathbf{H} \mathbf{A}_{\mathrm{cp}} \mathbf{Q}\\
\boldsymbol{\eta}_\mathrm{eff}&=&\mathbf{W}\boldsymbol{\eta}.
\end{IEEEeqnarray}
The covariance matrices of $\hat{\mathbf{s}}$ and $\boldsymbol{\eta}_\mathrm{eff}$ are given by
\begin{IEEEeqnarray}{rCL}
\mathbb{E}\left[\hat{\mathbf{s}}\hat{\mathbf{s}}^H\right]&=&\mathbb{E}\left[\left(\mathbf{H}_{\mathrm{eff}} \mathbf{x}+\boldsymbol{\eta}_\mathrm{eff}\right)\left(\mathbf{H}_{\mathrm{eff}} \mathbf{x}+\boldsymbol{\eta}_\mathrm{eff}\right)^H\right]\nonumber \\
&=&\sigma_x^2\mathbf{H}_\mathrm{eff} \mathbf{H}_\mathrm{eff} ^H+\sigma_0^2\mathbf{W} \mathbf{G}\mathbf{W}^H\\
\mathbb{E}\left[\boldsymbol{\eta}_\mathrm{eff}\boldsymbol{\eta}_\mathrm{eff}^H\right]&=&\sigma_0^2\mathbf{W} \mathbf{G}\mathbf{W}^H.
\end{IEEEeqnarray}

Hence, assuming that perfect CSI is available at the receiver, the information rate of the proposed scheme is given by~\cite{goldsmith2005wireless}
\begin{IEEEeqnarray}{rCL}
R&=&\log _2 \frac{\left|\sigma_x^2\mathbf{H}_\mathrm{eff} \mathbf{H}_\mathrm{eff} ^H+\sigma_0^2\mathbf{W} \mathbf{G}\mathbf{W}^H\right|_{\det}}{\left|\sigma_0^2\mathbf{W} \mathbf{G}\mathbf{W}^H\right|_{\det}}\nonumber \\
&=&\log _2\left|\mathbf{I}+\frac{\sigma_x^2}{\sigma_0^2} (\mathbf{W} \mathbf{G} \mathbf{W}^H)^{-1} \mathbf{H}_\mathrm{eff} \mathbf{H}_\mathrm{eff}^H\right|_{\det}\nonumber\\
&=&\log _2\left|\mathbf{I}+\frac{\sigma_x^2}{\sigma_0^2} \mathbf{G}_{\mathrm{z}}^{-1} \mathbf{H}_\mathrm{eff} \mathbf{H}_\mathrm{eff}^H\right|_{\det},\label{inf}
\end{IEEEeqnarray}
where $\mathbf{G}_{\mathrm{z}}=\mathbf{W} \mathbf{G}\mathbf{W}^H$. Since $\mathbf{G}_{\mathrm{z}}$ is a Hermitian matrix, i.e., $\mathbf{G}_{\mathrm{z}}^H=\mathbf{G}_{\mathrm{z}}$, the corresponding EVD is expressed as follows:
\begin{IEEEeqnarray}{rCL}
\mathbf{G}_{\mathrm{z}}=\mathbf{V} \boldsymbol{\Lambda} \mathbf{V}^{H}, \label{Gz}
\end{IEEEeqnarray}
\textcolor{black}{where $\boldsymbol{\Lambda}\in \mathbb{C}^{MN\times MN}$ is a diagonal matrix and $\mathbf{V} \in \mathbb{C}^{MN\times MN}$ is the unitary matrix.}
With the aid of \eqref{Gz}, \eqref{inf} is further simplified to
\begin{IEEEeqnarray}{rCL}
R&=&\log _2\left|\mathbf{I}+\frac{\sigma_x^2}{\sigma_0^2}  \mathbf{H}_\mathrm{eff}^H(\mathbf{W} \mathbf{G} \mathbf{W}^H)^{-1} \mathbf{H}_\mathrm{eff}\right|_{\det}\nonumber \\
&=&\log _2\left|\mathbf{I}+\frac{\sigma_x^2}{\sigma_0^2}  \mathbf{H}_\mathrm{eff}^H\mathbf{V} \boldsymbol{\Lambda}^{-\frac{1}{2}} \boldsymbol{\Lambda}^{-\frac{1}{2}} \mathbf{V}^{H} \mathbf{H}_\mathrm{eff}\right|_{\det}.\label{infevd}
\end{IEEEeqnarray}
Let us now define $\mathbf{D}_0=\boldsymbol{\Lambda}^{-\frac{1}{2}} \mathbf{V}^{H} \mathbf{H}_\mathrm{eff}$. Then the EVD of $\mathbf{D}_0^H \mathbf{D}_0$ is given by $\mathbf{D}_0^H \mathbf{D}_0=\mathbf{U}_0 \boldsymbol{\Xi} \mathbf{U}_0^H$, where
$\mathbf{U}_0 \in \mathbb{C}^{MN\times MN}$ represents a unitary matrix, \textcolor{black}{and the diagonal matrix $\boldsymbol{\Xi}\in \mathbb{C}^{MN\times MN}$ has $MN$ eigenvalues of $\xi_0, \cdots ,\xi_{MN-1}$, arranged in descending order.}
Hence, \eqref{infevd} is rewritten as
\begin{IEEEeqnarray}{rCL}
R&=&\log _2\left|\mathbf{I}+\frac{\sigma_x^2}{\sigma_0^2} \mathbf{D}_0^H \mathbf{D}_0\right|_{\det}\nonumber \\
\label{Rinf}&=&\sum_{i=0}^{MN-1} \log _2\left(1+\frac{\sigma_x^2 \xi_i}{\sigma_0^2} \right).
\end{IEEEeqnarray}

Upon normalizing \eqref{Rinf} by the bandwidth $2W(1+\beta)$ and the frame duration $MN\alpha T_0$, the achievable information rate is given by
\begin{IEEEeqnarray}{rCL}
R_N&=&\frac{1}{2W(1+\beta)MN\alpha T_0}\sum_{i=0}^{MN-1} \log _2\left(1+\frac{\sigma_x^2 \xi_i}{\sigma_0^2} \right). 
\end{IEEEeqnarray}

\section{Performance Results}
\label{simulation}
\begin{figure}[tbp]
\subfigure[]{
\centering
\includegraphics[width=0.43\linewidth]{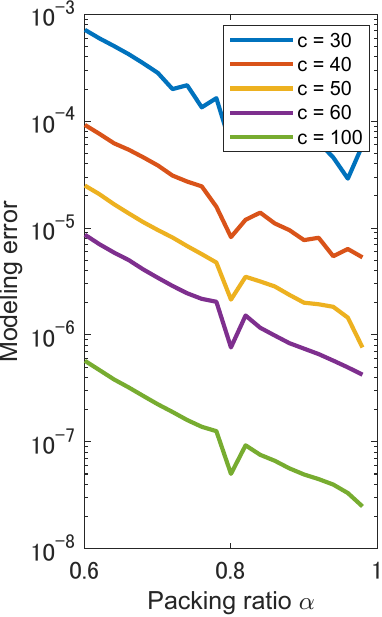}
}
\hspace{0.02\textwidth}
\subfigure[]{
\centering
\includegraphics[width=0.42\linewidth]{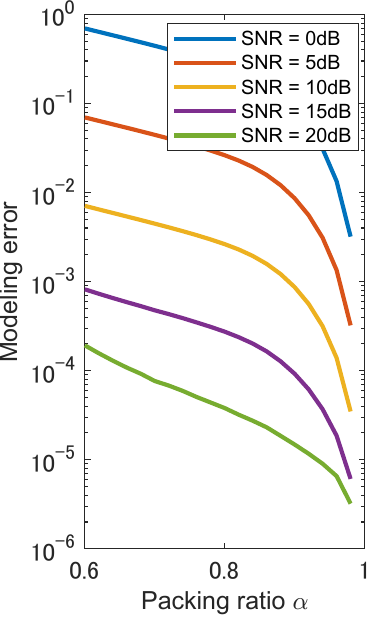}
}
\caption{Modeling errors of the channels and weights: (a) $\epsilon_0=\left\|\mathbf{H}_{\mathrm{t}}-\mathbf{H}_{\mathrm{s}}\right\|_F^2$, and (b) $\epsilon_1=\left\|\mathbf{W}_{\mathrm{t}}-\mathbf{W}_{\mathrm{1}}\right\|_F^2$.}
\label{modelerror}
\end{figure}
\textcolor{black}{In this section we provide performance results for characterizing the proposed OTFS-FTN scheme.}
We assume that each channel coefficient $h_i$ obeys the distribution of $h_i \sim \mathcal{C N}(0,1 /P)$ per frame.
\textcolor{black}{By using Jakes' model, each channel tap experiences a Doppler shift of $\nu_i=\nu_{\max } \cos \left(\theta_i\right)$, where $\theta_i$ is uniformly distributed over $[-\pi, \pi]$.} 
Moreover, the roll-off factor and the false alarm rate are set to $\beta=0.25$ and $P_{fa}=0.01$, respectively, unless otherwise stated.

In Fig.~\ref{modelerror}(a), we analyzed the modeling error $\epsilon_0=\left\|\mathbf{H}_{\mathrm{t}}-\mathbf{H}_{\mathrm{s}}\right\|_F^2$, induced due to the approximation of \eqref{aprHs}, where the packing ratio is varied from $\alpha=0.6$ to $0.98$ with the step size of $0.02$. 
\textcolor{black}{The maximum number of integer Doppler-shift taps is given by $k_{\max}=5$ and the maximum number of integer delay taps is $l_{\max}=15$.} Other parameters are set as $(M,N,P)=(128,12,15)$.
From Fig.~\ref{modelerror}(a), upon increasing the packing ratio or the CP length, the modeling errors are decreased. Next, Fig.~\ref{modelerror}(b) shows the modeling error $\epsilon_1=\left\|\mathbf{W}_{\mathrm{t}}-\mathbf{W}_{\mathrm{1}}\right\|_F^2$ of the LMMSE weights at $c=50$ for different SNRs, where $\mathbf{W}_{\mathrm{t}}=\mathbf{H_\mathrm{t} H_\mathrm{t}}^{H}+{\sigma_0^2} \mathbf{G}$. Observe from \eqref{W_1} that the modeling error is mainly induced by the circulant approximation of ${\sigma_0^2} \mathbf{G}_\mathrm{c}$. Hence, as seen in Fig.~\ref{modelerror}(b), upon decreasing the SNR, the modeling error decreases.
\begin{figure}
\centering
\includegraphics[width=1\linewidth]{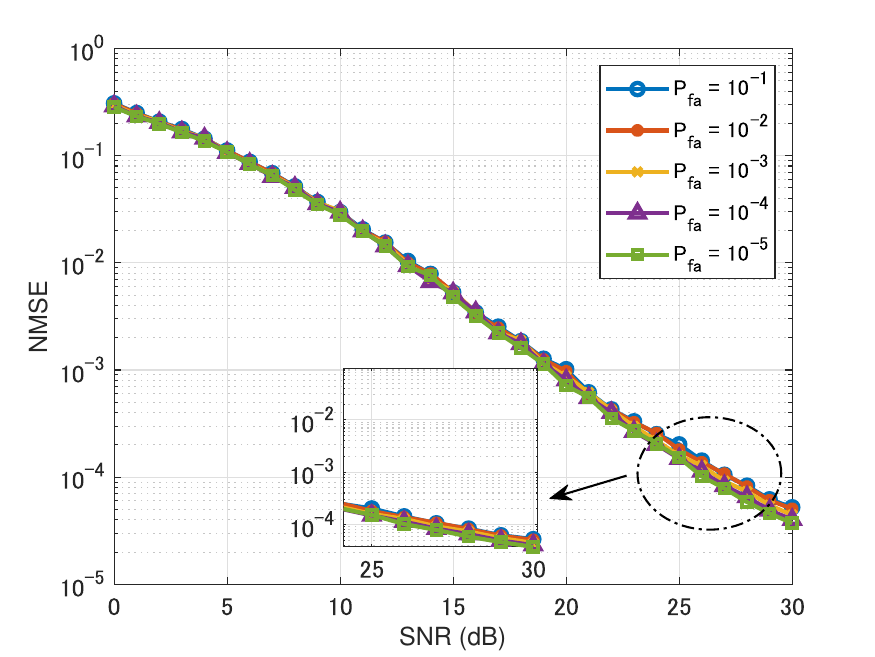}
\caption{NMSE of the proposed FTNP-based channel estimation for different false alarm rates.}
\label{NMSEfa}
\end{figure}
\begin{figure}
\centering
\includegraphics[width=1\linewidth]{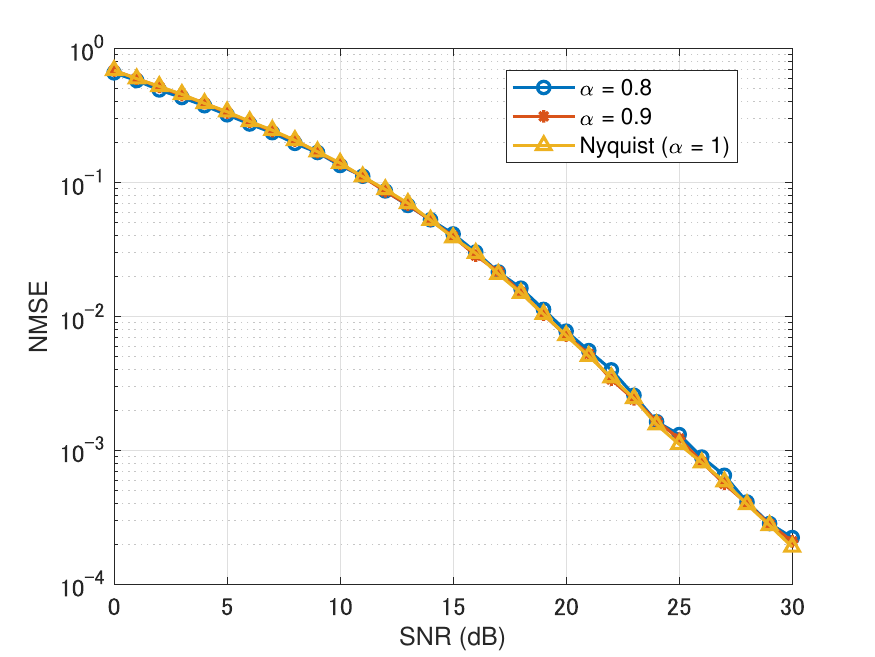}
\caption{NMSE of the proposed FTNP-based channel estimation for different packing ratios.}
\label{NMSE}
\end{figure}
\begin{figure}
\centering
\includegraphics[width=1\linewidth]{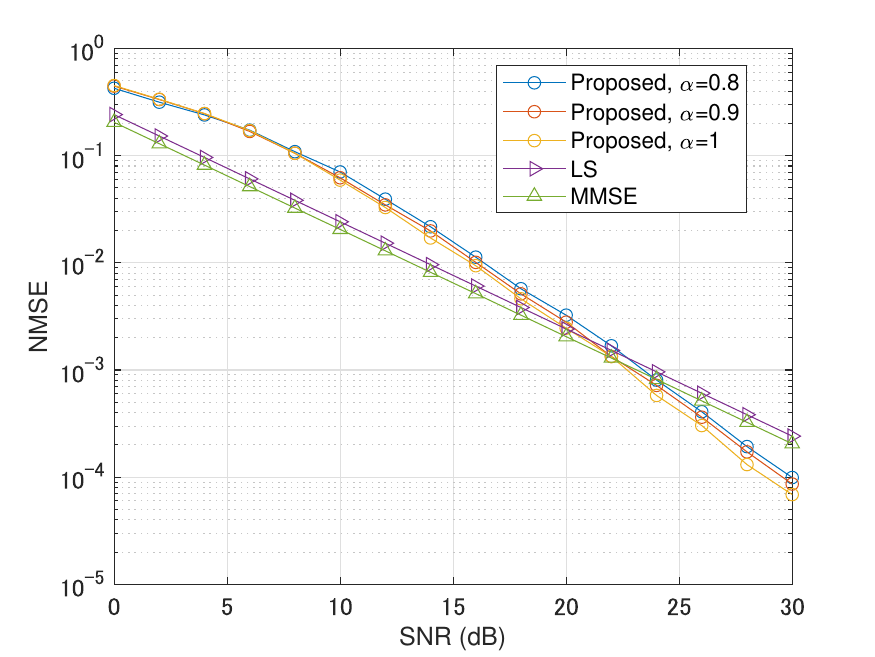}
\caption{\textcolor{black}{NMSE of the proposed FTNP-based channel estimation compared with other channel estimation schemes.}}
\label{NMSEcom}
\end{figure}
\begin{figure}[tbp]
\subfigure[]{
\centering
\includegraphics[width=0.43\linewidth]{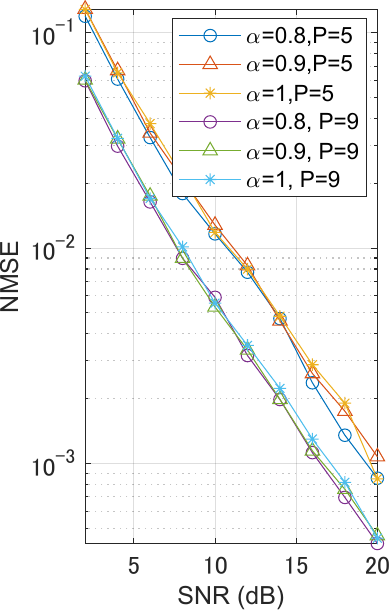}
}
\hspace{0.02\textwidth}
\subfigure[]{
\centering
\includegraphics[width=0.42\linewidth]{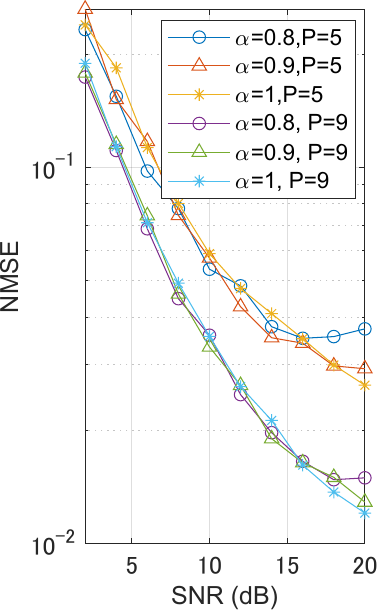}
}
\caption{\textcolor{black}{NMSE for (a) the distance estimation and (b) the velocity estimation.}}
\label{range_vel}
\end{figure}

Fig.~\ref{NMSEfa} shows the normalized mean square error (NMSE) of the proposed FTNP-based channel estimation scheme $\hat{\mathbf{H}}$ for different $P_{fa}$ in \eqref{T}. \textcolor{black}{The FTN packing ratio is fixed in order to isolate the impact of different false alarm rates on the estimation accuracy. The parameters are set to $\alpha=0.85$, $\Delta f =20$ kHz, $\nu_{\max}=10$ kHz, $l_{\max}=9$ and $(M, N, P, c) = (256, 6, 9, 60)$.} Observe in Fig.~\ref{NMSEfa} that upon reducing the false alarm rate $P_{fa}$, the NMSE improves. 
\textcolor{black}{This occurs because, at high $P_{fa}$ values, the false detection probability increases, making it more likely to mistakenly identify non-existent paths.}
\textcolor{black}{Furthermore, Fig.~\ref{NMSE} shows the NMSE, where we consider $\Delta f =30$ kHz, $\nu_{\max}=10$ kHz, $l_{\max}=30$ and $(M, N, P, c) = (128, 10, 30, 60)$ for different packing ratios.} It is found that our FTNP-based channel estimation exhibits an NMSE similar to the conventional Nyquist-based benchmark ($\alpha=1$).
\textcolor{black}{This behavior can be attributed to the FTN-aware DD-domain modeling, to the pilot-and-guard-based localized estimation strategy, and to the noise-whitening operation, which jointly mitigate the FTN-induced effects.}

\textcolor{black}{In Fig.~\ref{NMSEcom}, we compare the NMSE performance of the proposed FTNP-based channel estimation scheme to the Nyquist-based least squares (LS)~\cite{keykhosravi2023pilot} and MMSE estimators~\cite{ishihara2017iterative}.
We consider $\Delta f =30$ kHz, $\nu_{\max}=12$ kHz, $\tau_{\max}=2.51$ $\mu$s, and $(M, N, P, c,P_{fa}) = (256, 6, 9, 60,0.001)$.
It is observed that the proposed scheme achieves comparable or superior performance to the LS estimator and the MMSE estimator across the entire SNR range.
Furthermore, the proposed estimator remains robust under different FTN packing factors, including $\alpha=0.8$ and $\alpha=0.9$, indicating its efficiency in mitigating FTN-induced interference during the estimation process.
}

\textcolor{black}{Furthermore, in Fig.~\ref{range_vel}, we consider a monostatic radar sensing scenario based on the proposed OTFS-FTN signaling framework, where a single transceiver simultaneously performs transmission and echo reception.
Both the target range and the radial velocity are estimated from the delay and Doppler information extracted in the DD domain, similar to the scenario considered in~\cite{10713220}. 
Thus, the real-time distance $r_p$ and radial velocity $u_p$ of the $p$-th moving target are expressed as
\begin{IEEEeqnarray}{rCL}
r_p = \frac{c_l\tau_p}{2}, \qquad
u_p = \frac{c_l\nu_p}{f_c \cos\theta},
\end{IEEEeqnarray}
where $c_l$ denotes the speed of light and $f_c$ is the carrier frequency. Here, $\theta$ is the fixed azimuth angle of the target relative to the transmitter.
Accordingly, the distance and velocity of each target can be jointly estimated from the noisy DD-domain observations.
The NMSE of the estimated range and velocity is evaluated under different FTN packing factors and for different numbers of paths $P$ after reflection from the moving target. The other system parameters are the same as those used in Fig.~\ref{NMSEcom}.
The estimation accuracy improves monotonically upon increasing the SNR under all the configurations considered and for both metrics.
Compared to the Nyquist scheme ($\alpha=1$), FTN signaling exhibits a similar estimation performance, confirming its suitability for monostatic radar sensing in high-mobility environments.}
\begin{figure}
\centering
\includegraphics[width=1\linewidth]{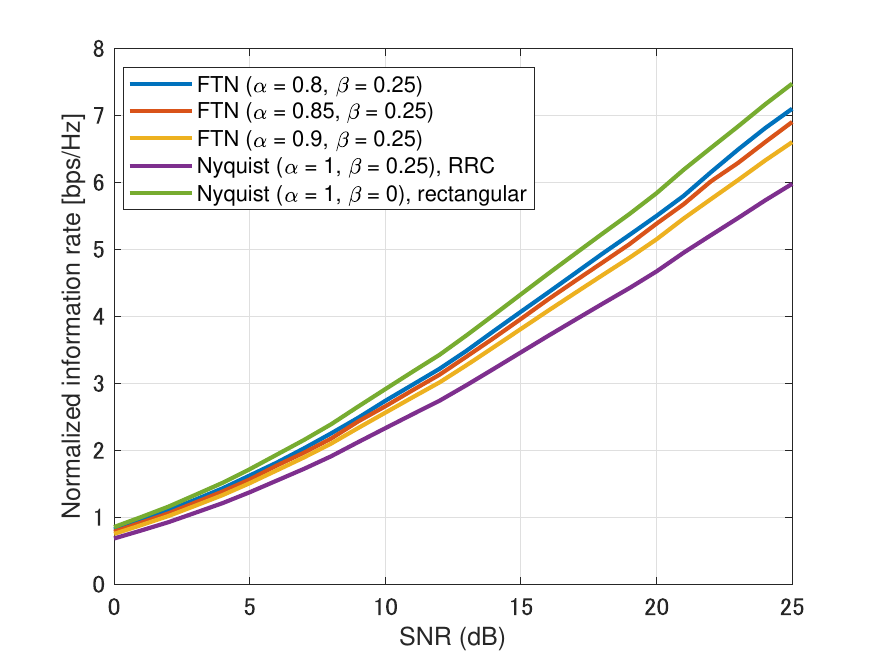}
\caption{\textcolor{black}{Achievable information rate of the proposed scheme under different packing ratios.}}
\label{Inf01}
\end{figure}
\begin{figure}
\centering
\includegraphics[width=1\linewidth]{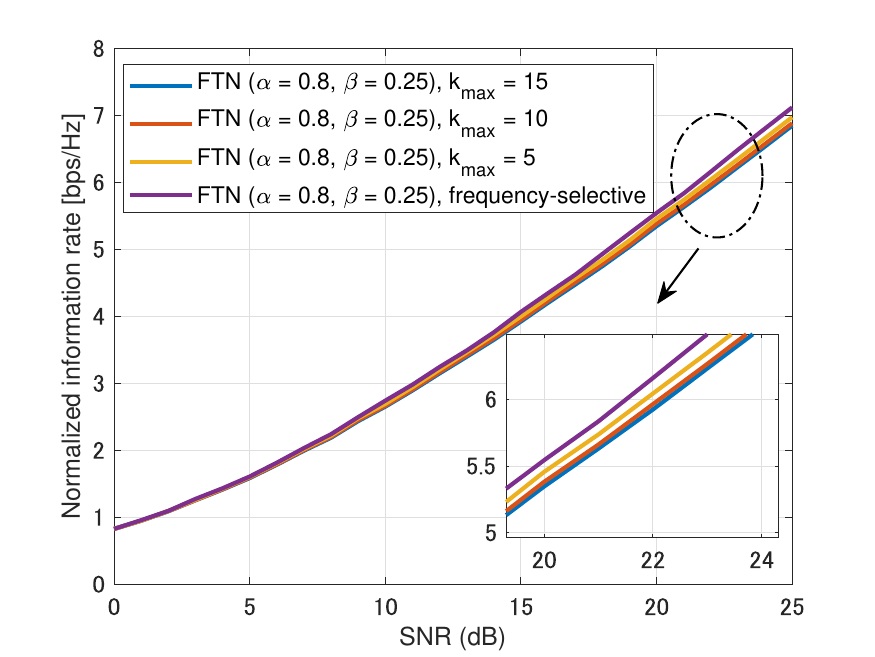}
\caption{\textcolor{black}{Achievable information rate of the proposed scheme under different Doppler shifts.}}
\label{Inf02}
\end{figure}
\begin{figure}
\centering
\includegraphics[width=1\linewidth]{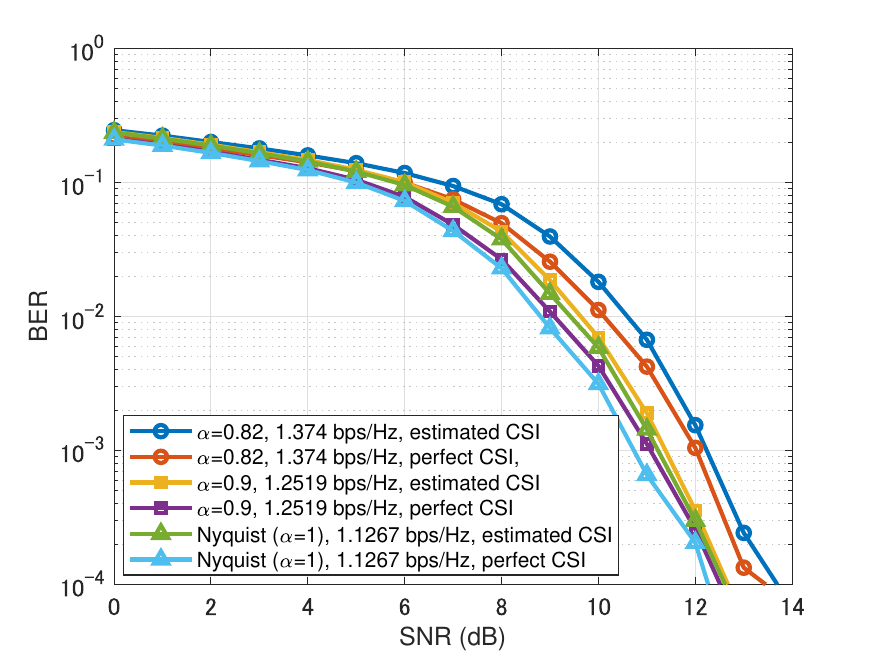}
\caption{\textcolor{black}{BER performance of the proposed scheme both with estimated and perfect CSI under different packing ratios.}}
\label{BER}
\end{figure}
\begin{figure}
\centering
\includegraphics[width=1\linewidth]{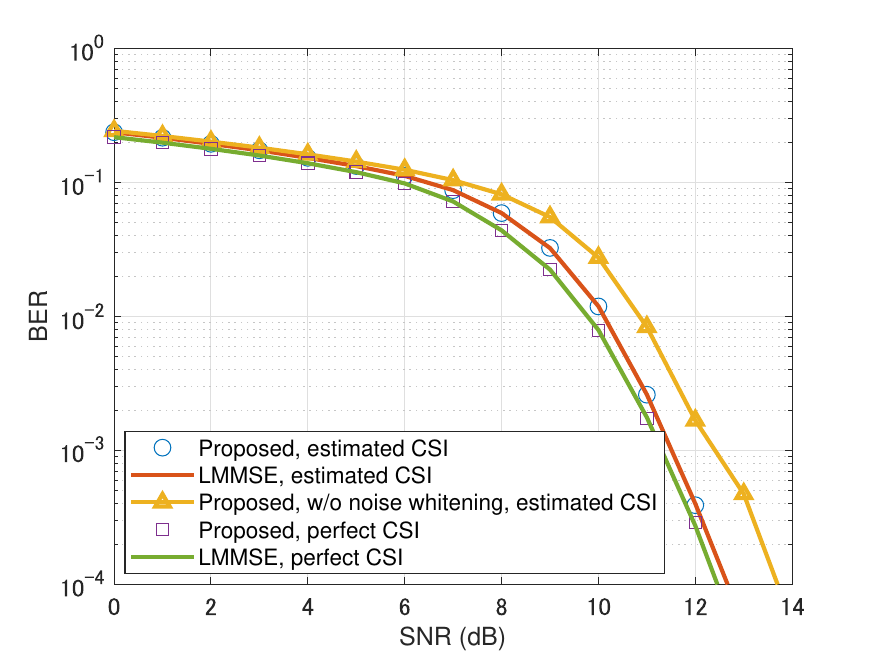}
\caption{BER performance of our approximated LMMSE equalizer with estimated and perfect CSI, compared with the full-complexity LMMSE bound of \eqref{zlmmse_act}.}
\label{BER_lmmse}
\end{figure}

Fig.~\ref{Inf01} shows the achievable information rate of the proposed OTFS-FTN signaling for different packing ratios. 
\textcolor{black}{Moreover, we also consider the conventional time-orthogonal OTFS benchmarks using the ideal rectangular shaping filter ($\beta=0$) and the RRC filter ($\beta=0.25$). 
The remaining system parameters are identical to those in Fig.~\ref{modelerror}.}
Observe in Fig.~\ref{Inf01} that the proposed OTFS-FTN signaling outperforms the Nyquist-based benchmark employing the RRC filter. Moreover, upon decreasing the packing ratio, the information rate increases while approaching the upper bound of the Nyquist-based OTFS scheme employing the rectangular shaping filter.
Furthermore, Fig.~\ref{Inf02} presents the achievable information rate of the proposed scheme for the maximum Doppler taps of $k_{\max}=5$, $10$, and $15$. \textcolor{black}{The other parameters are given by $(M,N,P,\alpha,l_{\max})=(64,20,30,0.8,30)$.}
\textcolor{black}{For comparisons, we also consider the proposed OTFS-FTN signaling under a time-invariant frequency-selective fading channel.
Observe from Fig.~\ref{Inf02} that as the maximum Doppler tap decreases, the information rate increases while approaching the performance of the time-invariant channel.}

\textcolor{black}{Next, we characterized the proposed scheme's BER performance. 
In this simulation a 3/4-rate low-density parity check (LDPC) coding scheme having a maximum of 50 iterations was considered. We define the transmission rate as follows:}
\begin{IEEEeqnarray}{rCL}
R_{\mathrm{t}}=\frac{3}{4} \cdot \frac{1}{2 W(1+\beta)} \cdot \frac{1}{MN \alpha T_0} \cdot \sum_{n=0}^{MN-1} b_n[\mathrm{bps} / \mathrm{Hz}],  \ \ \
\end{IEEEeqnarray}
\textcolor{black}{where $b_n$ represents the bits mapped to the $n$th symbol.}
Fig.~\ref{BER} \textcolor{black}{presents} the BER curves of the proposed OTFS-FTN scheme, employing  FTNP-based channel estimation for different packing ratios. For reference, the BER curves associated with perfect CSI are also plotted. \textcolor{black}{We consider QPSK, $\Delta f =30$ kHz, $\nu_{\max}=15$ kHz, $l_{\max}=15$ and $(M, N, P, c) = (256, 6, 15, 50)$.} Observe from Fig.~\ref{BER} that upon increasing the SNR, the BER gap between the estimated and perfect CSI decreases.

\textcolor{black}{Fig.~\ref{BER_lmmse} presents the BERs of the proposed OTFS-FTN scheme, employing our approximated LMMSE equalizer of \eqref{zlmmse_act}, where we consider $(M, N, P, c,\alpha) = (128, 12, 12, 50,0.83)$, $\Delta f =30$ kHz, $\nu_{\max}=10$ kHz, $l_{\max}=12$, and QPSK.} For comparison, we also plotted the BER curve of the proposed receiver without noise whitening.
It is seen in Fig.~\ref{BER_lmmse} that the proposed scheme achieved the BER performance of the full-complexity LMMSE bound, while outperforming the proposed scheme dispensing with noise whitening.

\begin{figure}
\centering
\includegraphics[width=1\linewidth]{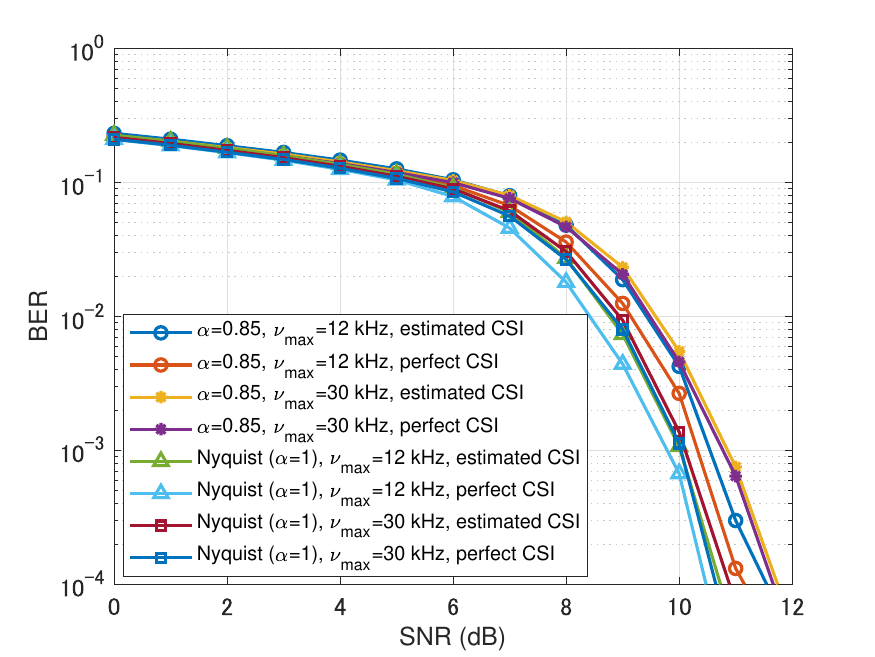}
\caption{BER performance of the proposed OTFS-FTN signaling scheme with estimated and perfect CSI for different maximum Doppler shifts.}
\label{BERdoppler}
\end{figure}

\textcolor{black}{Furthermore, in Fig.~\ref{BERdoppler} we compared the BERs of the proposed scheme under different maximum Doppler shifts, where the parameters are set as $(M, N, P, c) = (64, 30, 20, 60)$, $\Delta f =30$ kHz, $l_{\max}=20$, and QPSK.}
Observe in Fig.~\ref{BERdoppler} that despite varying the maximum Doppler shifts, the BER curves remained nearly unchanged, exhibiting robustness over the Doppler shift.
\begin{figure}
\centering
\includegraphics[width=1\linewidth]{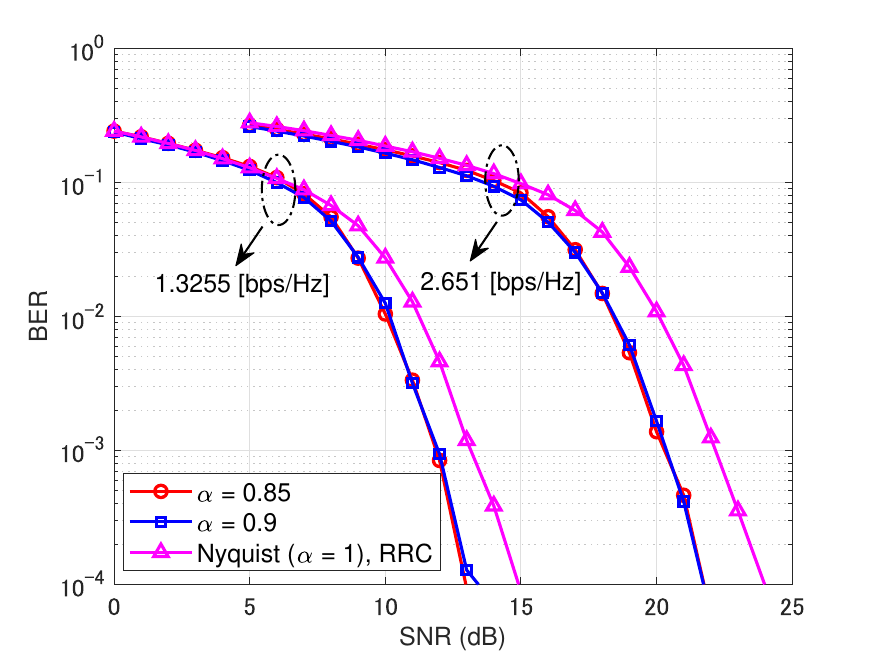}
\caption{BER performance of the proposed OTFS-FTN signaling scheme for different packing ratios.}
\label{BERtr}
\end{figure}
\begin{figure}
\centering
\includegraphics[width=1\linewidth]{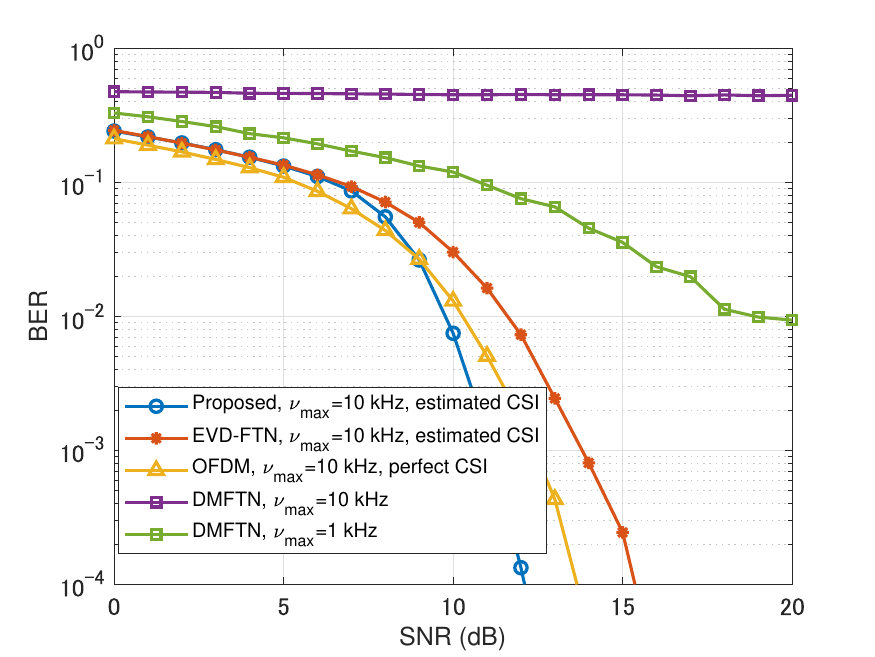}
\caption{BER performance of the proposed OTFS-FTN signaling scheme, compared with the conventional EVD-based FTN signaling benchmark~\cite{ishihara2021eigendecomposition}, DMFTN signaling \cite{ishihara2023differential}, and the conventional OFDM benchmark.}
\label{BER_compare}
\end{figure}
\begin{figure}
\subfigure[]{
\centering
\includegraphics[width=1\linewidth]{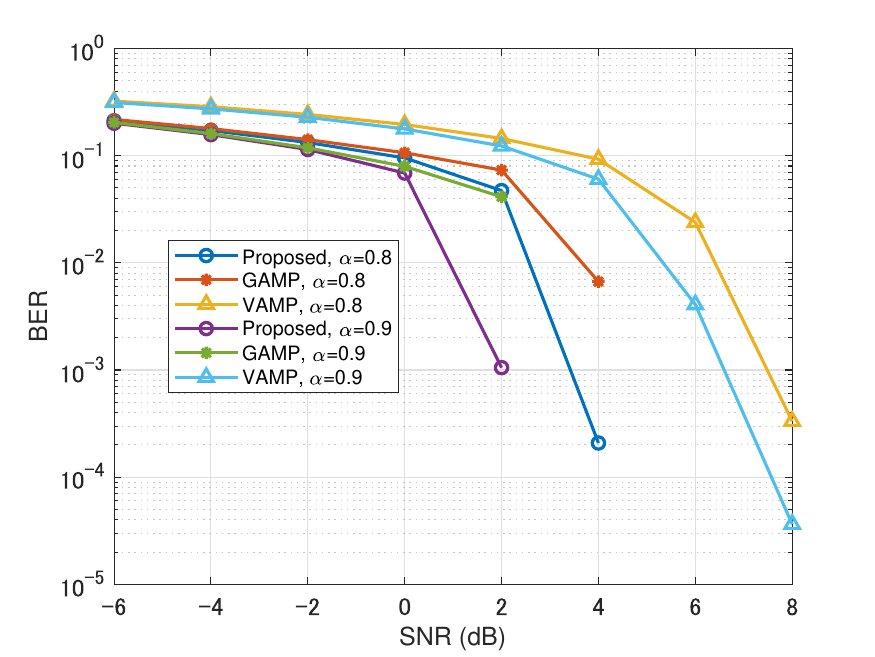}
}
\hspace{0.02\textwidth}
\subfigure[]{
\centering
\includegraphics[width=1\linewidth]{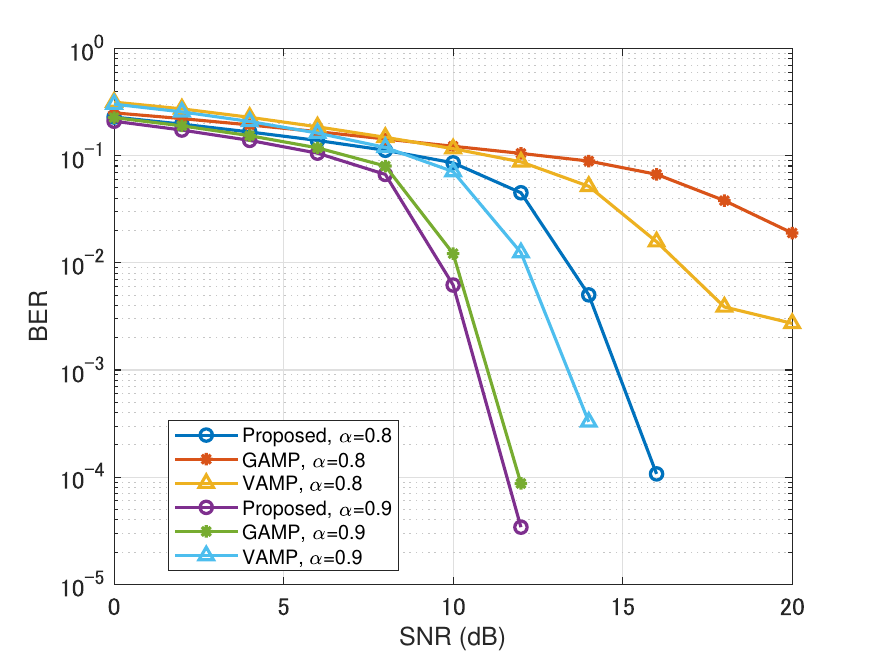}
}
\caption{\textcolor{black}{BER performance under EVA channel for OTFS-FTN systems compared with other detection schemes: (a) QPSK, and (b) 16QAM.}}
\label{BER_EVA}
\end{figure}
\begin{figure}
\centering
\includegraphics[width=1\linewidth]{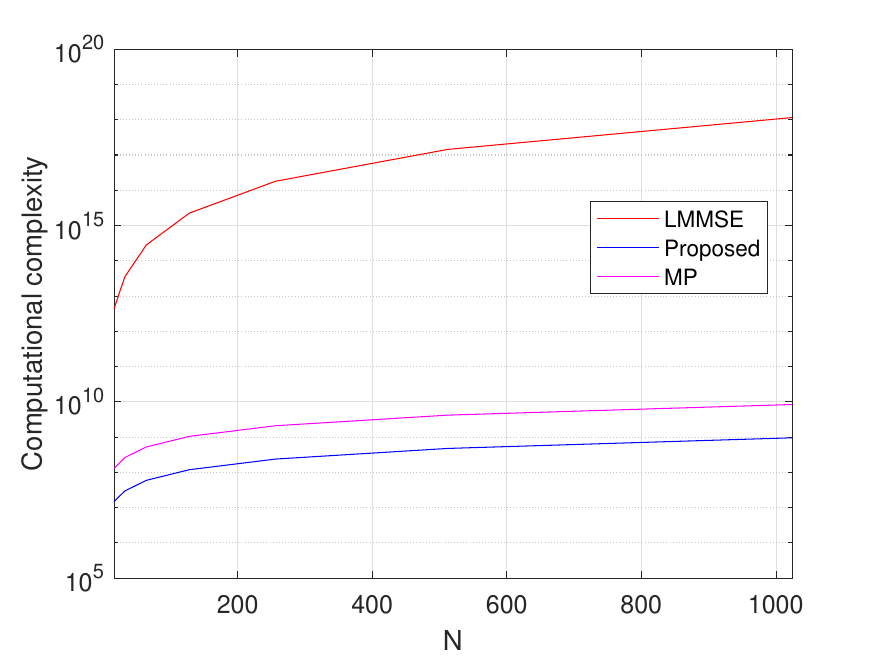}
\caption{Complexity comparison among the proposed detector, the LMMSE detector, and the MP detector.}
\label{CM_compare}
\end{figure}
\begin{figure}
\centering
\includegraphics[width=1\linewidth]{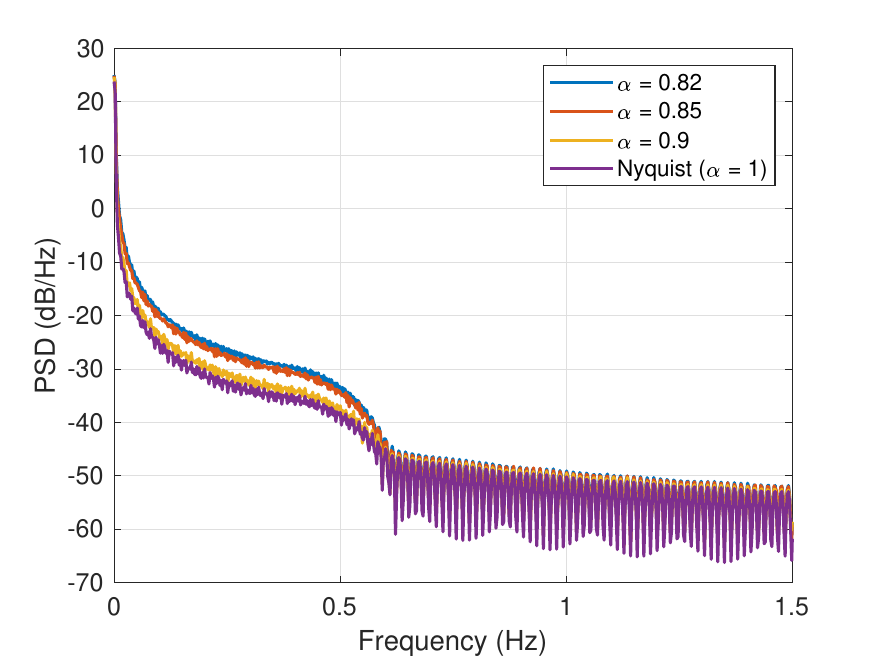}
\caption{PSD of the proposed OTFS-FTN signaling scheme for different packing ratios.}
\label{PSD}
\end{figure}

\textcolor{black}{In Fig.~\ref{BERtr}, we compared the BERs of the proposed OTFS-FTN signaling having the packing ratio of $\alpha =0.85$ and $0.9$ to the Nyquist-based OTFS benchmark, using the same RRC filter.}
The target transmission rate is set to $R_t=1.3255$ bps/Hz and $2.651$ bps/Hz. \textcolor{black}{The remaining parameters are given by $(M, N, P, c) = (128, 12, 10, 50)$,  $\nu_{\max}=7.5$ kHz, $l_{\max}=10$, and $\Delta f =30$ kHz.}
Here, to achieve the target transmission rate~$R_t$, we employ the bit-loading technique of~\cite{ishihara2019svd,ishihara2021eigendecomposition}, 
\textcolor{black}{where each symbol is assigned one of the following modulation schemes: QPSK, 16-QAM, or 64-QAM.} 
Observe in Fig.~\ref{BERtr} that the proposed scheme outperforms its Nyquist-based benchmark for each target rate scenario.
\textcolor{black}{More specifically, when the BER reaches $10^{-4}$, the proposed scheme achieves an approximately 2 dB performance gain over its Nyquist-based benchmark.
}

In Fig.~\ref{BER_compare}, we compared the BERs of the proposed scheme to those of the three benchmarks, i.e., the open-loop CSI-free differential multi-carrier FTN (DMFTN) signaling scheme~\cite{ishihara2023differential}, the equal-power EVD-based FTN (EVD-FTN) signaling scheme \cite{ishihara2021eigendecomposition}, and the conventional OFDM ($\alpha=1$) having perfect CSI.
\textcolor{black}{We consider QPSK, $\Delta f =20$ kHz, $\nu_{\max}=10$ kHz, $l_{\max}=15$, and $(M, N, P, c,\alpha) = (256, 6, 15, 60,0.85)$.}
For the EVD-FTN signaling benchmark, we consider the proposed FTNP-based channel estimation of the DD-domain to calculate the transmit precoding matrix by assuming channel reciprocity.
In Fig.~\ref{BER_compare}, it is found that even under channel estimation errors, the proposed scheme outperforms both the Nyquist-based OFDM scheme having perfect CSI and the other FTN benchmarks considered.

\textcolor{black}{In Figs.~\ref{BER_EVA}(a) and \ref{BER_EVA}(b), we compare the BER performance under the extended vehicular A (EVA) channel model~\cite{3gpp.36.101} for QPSK and 16QAM, respectively. 
Furthermore, the generalized approximate message passing (GAMP) detector of~\cite{ma2021generalized} and the vector approximate message passing (VAMP) detector of~\cite{ma2023vamp} are used as benchmarks, with the number of iterations set to 100.
The remaining parameters are $(M, N, P, c) = (128, 12, 9, 50)$,  $\nu_{\max}=12$ kHz, $\tau_{\max}=2.51$ $\mu s$ and $\Delta f =30$ kHz.
The proposed detector is capable of significantly outperforming the GAMP- and VAMP-based schemes in the moderate-to-high SNR region.
This performance advantage becomes more pronounced, when a more aggressive FTN packing factor ($\alpha=0.8$) is employed, indicating the robustness of the proposed LU-based equalizer against strong FTN-induced ISI.
For 16QAM, the performance degradation of GAMP and VAMP is further exacerbated by the sensitivity to model mismatch under high-mobility conditions. By contrast, the proposed method exhibits stable convergence and achieves reliable detection performance.}

\textcolor{black}{
Fig.~\ref{CM_compare} compares the complexity of the proposed detector, the conventional full-complexity LMMSE equalizer and the MP detector~\cite{raviteja2018interference}.
The computational complexity of the MP detector is $O[N_I MN P^2 Q_b]$,where $N_I$ is the number of iterations and $Q_b$ denotes the constellation order. We consider the value $N$ is varied and other parameters are $(c,N_I,P,M,Q_b)=(30,20,10,1024,4)$.
As shown in Fig.~\ref{CM_compare}, the conventional LMMSE equalizer exhibits a steep increase in complexity due to the complexity of $O\left[M^3 N^3\right]$.
In contrast, the proposed detector exhibits the lowest computational complexity among all considered detectors.
}

Finally, Fig.~\ref{PSD} shows the power spectral densities (PSD) of the proposed OTFS-FTN signaling scheme for different packing ratios. We consider $T_0=1$ for normalization, while employing BPSK and $(M,N,c)=(128,10,30)$. Furthermore, we consider a time interval of $[-100T_0,(MN+2c+100)T_0]$, where $s(t)$ is sampled at the interval of $0.2T_0$. \textcolor{black}{Observe from Fig.~\ref{PSD} that at $W(1+\beta)=0.625$ Hz, the PSD of the proposed OTFS-FTN signaling scheme sharply drops, although upon decreasing $\alpha$, the level of the spectral side-lobe slightly increases. Hence, the proposed OTFS-FTN signaling scheme does not result in substantial bandwidth broadening.}

\section{Conclusions}
\label{conclusion}
In this paper, we proposed DD-domain CSI estimation and reduced-complexity LMMSE detection for an OTFS-FTN signaling scheme communicating in doubly selective fading channels. 
\textcolor{black}{Based on the DD domain's input-output relationship of the OTFS-FTN scheme employing an RRC shaping filter, we designed efficient DD-domain FTNP-based channel estimation, achieving comparable performance to conventional Nyquist-based schemes.}
Furthermore, by utilizing a sparsity approximation of the FTN-induced ISI matrix, we proposed reduced-complexity LMMSE equalization operating without any substantial BER performance degradation.
\textcolor{black}{Our performance results demonstrated that, under the same RRC filter, the proposed OTFS-FTN signaling scheme achieved an improved BER performance to other conventional FTN schemes, while exhibiting a higher information rate than its conventional Nyquist-based OTFS counterpart.}
\textcolor{black}{Furthermore, when the BER reaches $10^{-4}$, the proposed scheme achieves approximately a 2 dB performance gain over its Nyquist-based OTFS benchmark at the same transmission rate.}
\appendices

\section{Proof of Theorem 1: OTFS-FTN Input-Output Relationship in Delay-Doppler Domain for RRC Pulses}
The cross ambiguity function between $h_{\mathrm{tx}}(t)$ and $h_{\mathrm{rx}}(t)$ is defined as
\begin{IEEEeqnarray}{rCL}
A_{h_{\mathrm{rx}}, h_{\mathrm{tx}}}(t, f) \!\!\triangleq \!\!\int h_{\mathrm{rx}}^*\left(t^{\prime}-t\right) h_{\mathrm{tx}}\left(t^{\prime}\right) e^{-j 2 \pi f\left(t^{\prime}-t\right)} d t^{\prime}.
\end{IEEEeqnarray}
Let us assume that the term $H_{n, m}\left[n^{\prime}, m^{\prime}\right]$ includes the effects of transmit pulse, receive pulse, and channel fading as follows: \cite{raviteja2018interference}
\begin{IEEEeqnarray}{rCL}
&&H_{n, m}\left[n^{\prime}, m^{\prime}\right] =\nonumber\\
&&\iint h(\tau, \nu) A_{h_{\mathrm{rx}}, h_{\mathrm{tx}}}\left(\left(n-n^{\prime}\right) T-\tau,\left(m-m^{\prime}\right) \Delta f-\nu\right) \nonumber \\
&&\times e^{j 2 \pi\left(\nu+m^{\prime} \Delta f\right)\left(\left(n-n^{\prime}\right) T-\tau\right)} e^{j 2 \pi \nu n^{\prime} T} d \tau d \nu.\label{Hnm}
\end{IEEEeqnarray}
For $|\tau|<\tau_{\max }$ and $|\nu|<\nu_{\max }$, $H_{n, m}\left[n^{\prime}, m^{\prime}\right]$ is only becomes non-zero when $n^{\prime}=n-1$ or $n^{\prime}=n$.
Let us denote the received and transmitted symbols in the time-frequency domain as $\bar{\mathbf{Y}}$ and $\bar{\mathbf{X}}$, respectively.
Then, the time-frequency relationship can be written as \cite{raviteja2018interference}
\begin{IEEEeqnarray}{rCL}
\bar{Y}[n, m]&= & \sum_{n^{\prime}=n-1}^n \sum_{m^{\prime}=0}^{M-1} H_{n, m}\left[n^{\prime}, m^{\prime}\right] \bar{X}\left[n^{\prime}, m^{\prime}\right] \nonumber \\
\nonumber &= & H_{n, m}[n, m] \bar{X}[n, m] \\
\nonumber &&+\underbrace{\sum_{m^{\prime}=0, m^{\prime} \neq m}^{M-1} H_{n, m}\left[n, m^{\prime}\right] \bar{X}\left[n, m^{\prime}\right]}_{\text{ICI}} \\
&& +\underbrace{\sum_{m^{\prime}=0}^{M-1} H_{n, m}\left[n-1, m^{\prime}\right] \bar{X}\left[n-1, m^{\prime}\right]}_{\text{ISI}}, \label{tfinput}
\end{IEEEeqnarray}
where $\bar{Y}[n, m]$ and $\bar{X}[n^{\prime}, m^{\prime}]$ are the $n$th-row and $m$th column entries of $\bar{\mathbf{Y}}$ and $\bar{\mathbf{X}}$, respectively.
The second and third terms in \eqref{tfinput} correspond to the ICI and ISI, which are affected by the cross ambiguity function in $H_{n, m}\left[n^{\prime}, m^{\prime}\right]$.
Hence, for the ICI term, let the cross ambiguity function in $H_{n, m}\left[n, m^{\prime}\right]$ be denoted as $A_{c}=A_{h_{\mathrm{rx},}, h_{\mathrm{tx}}}\left( -\tau,\left(m-m^{\prime}\right) \Delta f-\nu\right)$.
If we assume $g(kT_{\mathrm{f}}) \approx 0$ for $|k|>M>c$ with the the sampling interval of $T_{\mathrm{f}}=T/M$, then $A_{c}$ can be simplified to
\begin{IEEEeqnarray}{rCL}
A_{c} &= &\!\int h_{\mathrm{rx}}^*\left(t^{\prime}+\tau_i\right) \! h_{\mathrm{tx}}\left(t^{\prime}\right) e^{-j 2 \pi\left(\left(m-m^{\prime}\right) \Delta f-\nu_i\right)\left(t^{\prime}+\tau_i\right)} d t^{\prime} \nonumber \\
&\approx&\! \frac{T}{M} \! \sum_{p=0}^{M-1-l_{i}} \! g(p T_{\mathrm{f}})e^{-j 2 \pi\left(\left(m-m^{\prime}\right) \Delta f-\nu_i\right)\left(\frac{p}{M \Delta f}+\tau_i\right)}\!.\label{Aici}
\end{IEEEeqnarray}
Similarly, for the ISI term in \eqref{tfinput}, the cross ambiguity function $A_{s}=A_{h_{\mathrm{rx},}, h_{\mathrm{tx}}}\left( T-\tau,\left(m-m^{\prime}\right) \Delta f-\nu\right)$ of $H_{n, m}\left[n-1, m^{\prime}\right]$ can be formulated as:
\begin{IEEEeqnarray}{rCL}
A_{s} \!\!&=&\!\!\int h_{\mathrm{rx}}^*\left(t^{\prime}\!-\!(T\!-\!\tau_i)\right) \! h_{\mathrm{tx}}\!\!\left(t^{\prime}\!\right)\!\! e^{-j 2 \pi\left(\left(m-m^{\prime}\right) \Delta f-\nu_i\right)\left(t^{\prime}+\tau_i-T\right)} \!d t^{\prime} \nonumber \\
&\approx& \!\frac{T}{M} \! \!\sum_{p=M-l_{i}}^{M-1} \! g(p T_{\mathrm{f}})e^{-j 2 \pi\left(\left(m-m^{\prime}\right) \Delta f-\nu_i\right)\left(\frac{p}{M \Delta f}+\tau_i-T\right)}.
\end{IEEEeqnarray}

Based on \eqref{tfinput} and with the SFFT, the symbols $\widetilde{Y}[k, l]$ received in the DD domain are represented by
\begin{IEEEeqnarray}{rCL}
\widetilde{Y}[k, l]&=&\frac{1}{\sqrt{N M}} \sum_{n=0}^{N-1} \sum_{m=0}^{M-1}\left(\sum_{m^{\prime}=0}^{M-1} H_{n, m}\left[n, m^{\prime}\right] \bar{X}\left[n, m^{\prime}\right]+\right.\nonumber\\
&&\left.\sum_{m^{\prime}=0}^{M-1} H_{n, m}\left[n-1, m^{\prime}\right] \bar{X}\left[n-1, m^{\prime}\right]\right) e^{-j 2 \pi\left(\frac{n k}{N}-\frac{m l}{M}\right)}\nonumber \\
\label{y_DD}&=&\widetilde{Y}_{c}[k, l]+\widetilde{Y}_{s}[k, l],
\end{IEEEeqnarray}
where we have:
\begin{IEEEeqnarray}{rCL}
\widetilde{Y}_{c}[k, l]&=&\frac{1}{\sqrt{N M}} \!\sum_{n=0}^{N-1}\! \sum_{m=0}^{M-1}\!\sum_{m^{\prime}=0}^{M-1} \!H_{n, m}\!\left[n, m^{\prime}\right] \bar{X}\left[n, m^{\prime}\right]\! \nonumber\\
&&\times e^{-j 2 \pi\left(\frac{n k}{N}-\frac{m l}{M}\right)} \\
\widetilde{Y}_{s}[k, l]&=&\frac{1}{\sqrt{N M}} \!\sum_{n=0}^{N-1}\! \sum_{m=0}^{M-1}\!\sum_{m^{\prime}=0}^{M-1} \!H_{n, m}\!\left[n-1, m^{\prime}\right] \bar{X}\left[n-1, m^{\prime}\right]\! \nonumber \\ &&\times e^{-j 2 \pi\left(\frac{n k}{N}-\frac{m l}{M}\right)}.
\end{IEEEeqnarray}

Using the ISFFT, $\widetilde{Y}_{c}[k, l]$ is rewritten as:
\begin{IEEEeqnarray}{rCL}
\widetilde{Y}_{c}[k, l]&=&\frac{1}{N M} \sum_{k^{\prime}=0}^{N-1} \sum_{l^{\prime}=0}^{M-1} \widetilde{X}\left[k^{\prime}, l^{\prime}\right]\sum_{n=0}^{N-1} \sum_{m=0}^{M-1} \sum_{m^{\prime}=0}^{M-1} H_{n, m}\left[n, m^{\prime}\right] \nonumber \\
&&\times e^{-j 2 \pi n\left(\frac{k-k^{\prime}}{N}\right)} e^{j 2 \pi\left(\frac{m l-m^{\prime} l^{\prime}}{M}\right)}\nonumber \\
&=&\frac{1}{N M} \sum_{k^{\prime}=0}^{N-1} \sum_{l^{\prime}=0}^{M-1} \widetilde{X}\left[k^{\prime}, l^{\prime}\right] h_c[k^{\prime},l^{\prime}],\label{yckl}
\end{IEEEeqnarray}
where
\begin{IEEEeqnarray}{rCL}
h_c[k^{\prime},l^{\prime}]\!&\!=\!\!&\!\!\sum_{n=0}^{N-1} \!\!\sum_{m=0}^{M-1} \!\sum_{m^{\prime}=0}^{M-1}\! \!\!H_{n, m}\!\left[n, m^{\prime}\right]\!e^{-j 2 \pi n\left(\frac{k-k^{\prime}}{N}\right)} \!e^{j 2 \pi\left(\frac{m l-m^{\prime} l^{\prime}}{M}\right)}.\nonumber\\ \label{eq_hc}
\end{IEEEeqnarray}
Based on \eqref{Hnm} and \eqref{Aici}, the terms related to $m$, $m^{\prime}$, $n$ and $p$ in $h_c[k^{\prime},l^{\prime}]$ of \eqref{eq_hc} are separated as follows:
\begin{IEEEeqnarray}{rCL}
h_c[k^{\prime},l^{\prime}]&\approx&\sum_{i=0}^{P-1} h_i\left[\sum_{n=0}^{N-1} e^{-j 2 \pi n\left(\frac{k-k^{\prime}-k_{i}-\kappa_{i}}{N}\right)}\right]\nonumber \\
&&\times \left[\frac{T}{M} \sum_{p=0}^{M-1-l_{i}} g(pT_{\mathrm{f}})e^{j 2 \pi \frac{p}{M}\left(\frac{k_{i}+\kappa_{i}}{N}\right)}  \right.\nonumber\\
&&\left.\times \sum_{m=0}^{M-1} e^{-j 2 \pi\left(p+l_{i}-l\right) \frac{m}{M}}\sum_{m^{\prime}=0}^{M-1} e^{j 2 \pi\left(p-l^{\prime}\right) \frac{m^{\prime}}{M}}\right]\nonumber \\
&=&\sum_{i=0}^{P-1} h_i \mathcal{F}_c \mathcal{G}_c, \label{hckl}
\end{IEEEeqnarray}
where we have:
\begin{IEEEeqnarray}{rCL}
\mathcal{F}_c&=&\sum_{n=0}^{N-1} e^{-j 2 \pi n\left(\frac{k-k^{\prime}-k_{i}-\kappa_{i}}{N}\right)}\nonumber \\
&=&\frac{e^{-j 2 \pi \left(k-k^{\prime}-k_{i}-\kappa_{i}\right)}-1}{e^{-j 2 \pi (\frac{k-k^{\prime}-k_{i}-\kappa_{i}}{N})}-1} \label{fc}
\end{IEEEeqnarray}
\begin{IEEEeqnarray}{rCL}
\mathcal{G}_c&=&\frac{T}{M} \sum_{p=0}^{M-1-l_{i}} g(pT_{\mathrm{f}})e^{j 2 \pi \frac{p}{M}\left(\frac{k_{i}+\kappa_{i}}{N}\right)} \sum_{m=0}^{M-1} e^{-j 2 \pi\left(p+l_{i}-l\right) \frac{m}{M}} \nonumber \\
&&\times \sum_{m^{\prime}=0}^{M-1} e^{j 2 \pi\left(p-l^{\prime}\right) \frac{m^{\prime}}{M}}. \label{gc}
\end{IEEEeqnarray}

Similar to the analysis of conventional OTFS having an ideal waveform~\cite{raviteja2018interference}, \eqref{gc} is further simplified to
\begin{IEEEeqnarray}{rCL}
\mathcal{G}_c\!\!&=&\!\!MT \!\!\sum_{p=0}^{M-1-l_{i}}\!\! g(pT_{\mathrm{f}})e^{j 2 \pi \frac{p}{M}\left(\frac{k_{i}+\kappa_{i}}{N}\right)}\! \delta\left(\left[p\!+\!l_{i}\!-\!l\right]_M\right) \!\delta\left(\!\left[p\!-\!l^{\prime}\right]_M\!\right).\nonumber \\ \label{gcs}
\end{IEEEeqnarray}
Hence, by substituting \eqref{hckl}, \eqref{fc} and \eqref{gcs} into \eqref{yckl}, we obtain
\begin{IEEEeqnarray}{rCL}
\widetilde{Y}_{c}[k, l]&\approx&\frac{T}{N} \sum_{i=0}^{P-1} h_i\left(\sum_{l^{\prime}=0}^{M-1} \sum_{p=0}^{M-1-l_{i}} g(pT_{\mathrm{f}})e^{j 2 \pi \frac{p}{M}\left(\frac{k_{i}+\kappa_{i}}{N}\right)}\right.\nonumber\\
&&\times \delta\left(\left[p+l_{i}-l\right]_M\right) \delta\left(\left[p-l^{\prime}\right]_M\right)\nonumber\\
&&\left.\times \sum_{k^{\prime}=0}^{N-1} \frac{e^{-j 2 \pi \left(k-k^{\prime}-k_{i}-\kappa_{i}\right)}-1}{e^{-j 2 \pi (\frac{k-k^{\prime}-k_{i}-\kappa_{i}}{N})}-1}\widetilde{X}\left[k^{\prime}, l^{\prime}\right]\right)\nonumber \\
&\approx&\frac{T}{N} \sum_{i=0}^{P-1} h_i\left(\sum_{p=0}^{M-1-l_{i}} g(pT_{\mathrm{f}})e^{j 2 \pi \frac{p}{M}\left(\frac{k_{i}+\kappa_{i}}{N}\right)}\right.\nonumber\\
&&\times \delta\left(\left[p+l_{i}-l\right]_M\right)\sum_{q=-N_i}^{N_i} \frac{e^{-j 2 \pi \left(-q-\kappa_{i}\right)}-1}{e^{-j 2 \pi (\frac{-q-\kappa_{i}}{N})}-1} \nonumber\\
&&\left.\times \widetilde{X}\left[[k-k_{i}+q]_{N}, p\right]\right),\label{ycklapr}
\end{IEEEeqnarray}
where $\left[k-k_{i}-N_i\right]_N \leq k^{\prime} \leq\left[k-k_{i}+N_i\right]_{N}$ and $N_i \ll N$, similar to~\cite{raviteja2018interference}.
From \eqref{ycklapr}, $\widetilde{Y}_{c}[k, l]$ is non-zero only if $p=l-l_{i}$ and $l \geq l_{i}$. Therefore, for $l \geq l_{i}$, $\widetilde{Y}_{c}[k, l]$ is expressed by
\begin{IEEEeqnarray}{rCL}
\widetilde{Y}_{c}[k, l]&\approx &\sum_{i=0}^{P-1} \sum_{q=-N_i}^{N_i} h_i e^{j 2 \pi \frac{\left(l-l_i\right)\left(k_i+\kappa_i\right)}{M N}}\! \frac{T}{N}\!g((l-l_i)T_{\mathrm{f}})\rho(q,\kappa_i)\nonumber\\
&&\times \widetilde{X}\left[\left[k-k_i+q\right]_N,\left[l-l_i\right]_M\right], \label{ycf}
\end{IEEEeqnarray}
noting that $\rho(q,\kappa_i)$ is defined in \eqref{DD3}.

Similar to the above-mentioned transformation of $\widetilde{Y}_{c}[k, l]$, $\widetilde{Y}_{s}[k, l]$ in \eqref{y_DD} is also expressed by
\begin{IEEEeqnarray}{rCL}
\widetilde{Y}_{s}[k, l]&=&\frac{1}{N M} \sum_{k^{\prime}=0}^{N-1} \sum_{l^{\prime}=0}^{M-1}e^{-j 2 \pi \frac{k^{\prime}}{N}} \widetilde{X}\left[k^{\prime}, l^{\prime}\right] \nonumber \\
&&\times\sum_{n=0}^{N-1} \sum_{m=0}^{M-1} \sum_{m^{\prime}=0}^{M-1} H_{n, m}\left[n-1, m^{\prime}\right] e^{-j 2 \pi n\left(\frac{k-k^{\prime}}{N}\right)} \nonumber \\
&&\times e^{j 2 \pi\left(\frac{m l-m^{\prime} l^{\prime}}{M}\right)} \nonumber \\
&=&\frac{1}{N M} \sum_{k^{\prime}=0}^{N-1} \sum_{l^{\prime}=0}^{M-1}e^{-j 2 \pi \frac{k^{\prime}}{N}} \widetilde{X}\left[k^{\prime}, l^{\prime}\right] h_s[k^{\prime},l^{\prime}],\label{yskl}
\end{IEEEeqnarray}
where we have
\begin{IEEEeqnarray}{rCL}
h_s[k^{\prime},l^{\prime}]&\approx&\sum_{i=0}^{P-1} h_i\left[\sum_{n=1}^{N-1} e^{-j 2 \pi n\left(\frac{k-k^{\prime}-k_{i}-\kappa_{i}}{N}\right)}\right]\nonumber \\
&&\times \left[\frac{T}{M} \sum_{p=M-l_i}^{M-1} g(pT_{\mathrm{f}})e^{j 2 \pi (\frac{p-M}{M})\left(\frac{k_{i}+\kappa_{i}}{N}\right)}  \right.\nonumber\\
&&\left.\times \sum_{m=0}^{M-1} e^{-j 2 \pi\left(p+l_{i}-l-M\right) \frac{m}{M}}\sum_{m^{\prime}=0}^{M-1} e^{j 2 \pi\left(p-l^{\prime}\right) \frac{m^{\prime}}{M}}\right]\nonumber \\
&=&\sum_{i=0}^{P-1} h_i \mathcal{F}_s \mathcal{G}_s, \label{hskl}
\end{IEEEeqnarray}
with
\begin{IEEEeqnarray}{rCL}
\mathcal{F}_s&=&\sum_{n=1}^{N-1} e^{-j 2 \pi n\left(\frac{k-k^{\prime}-k_{i}-\kappa_{i}}{N}\right)}\nonumber\\
&=&\frac{e^{-j 2 \pi \left(k-k^{\prime}-k_{i}-\kappa_{i}\right)}-1}{e^{-j 2 \pi (\frac{k-k^{\prime}-k_{i}-\kappa_{i}}{N})}-1}-1\label{F_s}
\end{IEEEeqnarray}
\begin{IEEEeqnarray}{rCL}
\mathcal{G}_s&=&\frac{T}{M} \sum_{p=M-l_i}^{M-1} g(pT_{\mathrm{f}})e^{j 2 \pi (\frac{p-M}{M})\left(\frac{k_{i}+\kappa_{i}}{N}\right)} \!\!\nonumber\\
&&\times \sum_{m=0}^{M-1} e^{-j 2 \pi\left(p+l_{i}-l-M\right) \frac{m}{M}}\sum_{m^{\prime}=0}^{M-1} e^{j 2 \pi\left(p-l^{\prime}\right) \frac{m^{\prime}}{M}}\!\! \nonumber \\
&=&\!\!M T\!\!\!\!\sum_{p=M-l_i}^{M-1}\!\!\!\!\!g(pT_{\mathrm{f}})e^{j 2 \pi (\frac{p-M}{M})\!\left(\frac{k_{i}+\kappa_{i}}{N}\right)}\! \delta\!\left(\left[p\!+\!l_{i}\!-\!l\right]_M\right) \!\delta\!\left(\!\left[p\!-\!l^{\prime}\right]_M\!\right).\!\!\nonumber \\ \label{G_s}
\end{IEEEeqnarray}
By substituting \eqref{hskl}, \eqref{F_s}, and \eqref{G_s} into \eqref{yskl}, we can obtain
\begin{IEEEeqnarray}{rCL}
\widetilde{Y}_{s}[k, l]\!&\approx&\!\frac{T}{N} \sum_{i=0}^{P-1} h_i\left(\!\sum_{l^{\prime}=0}^{M-1} \sum_{p=M-l_i}^{M-1} g(pT_{\mathrm{f}})e^{j 2 \pi (\frac{p-M}{M})\left(\frac{k_{i}+\kappa_{i}}{N}\!\right)}\right.\nonumber\\
&&\times \delta\left(\left[p+l_{i}-l\right]_M\right) \delta\left(\left[p-l^{\prime}\right]_M\right)\nonumber\\
&&\left.\times \sum_{k^{\prime}=0}^{N-1} (\frac{e^{-j 2 \pi \left(k-k^{\prime}-k_{i}-\kappa_{i}\right)}\!-\!1}{e^{-j 2 \pi (\frac{k-k^{\prime}-k_{i}-\kappa_{i}}{N})}\!-\!1}\!-\!1)e^{-j 2 \pi \frac{k^{\prime}}{N}}\widetilde{X}\!\left[k^{\prime}, l^{\prime}\right]\right)\nonumber \\
&\approx&\frac{T}{N} \sum_{i=0}^{P-1} h_i\left(\sum_{p=M-l_i}^{M-1} g(pT_{\mathrm{f}})e^{j 2 \pi (\frac{p-M}{M})\left(\frac{k_{i}+\kappa_{i}}{N}\right)}\right.\nonumber\\
&&\times \delta\left(\left[p+l_{i}-l\right]_M\right)\!\sum_{q=-N_i}^{N_i} \!\left(\frac{e^{-j 2 \pi \left(-q-\kappa_{i}\right)}-1}{e^{-j 2 \pi (\frac{-q-\kappa_{i}}{N})}-1}-1\right) \nonumber\\
&&\left.\times e^{-j 2 \pi \frac{\left[k-k_{i}+q\right]_N}{N}} \widetilde{X}\left[[k-k_{i}+q]_{N}, p\right]\right).
\end{IEEEeqnarray}
Similar to \eqref{ycklapr}, $\widetilde{Y}_{s}[k, l]$ is non-zero only if $p=l-l_{i}+M$ and $l < l_{i}$. Hence, for $l <l_{i}$, $\widetilde{Y}_{s}[k, l]$ is approximated as:
\begin{IEEEeqnarray}{rCL}
&&\widetilde{Y}_{s}[k, l]\approx \sum_{i=0}^{P-1} \sum_{q=-N_i}^{N_i} h_i e^{j 2 \pi \frac{\left(l-l_i\right)\left(k_i+\kappa_i\right)}{M N}} \!\frac{T}{N}\!g((l-l_i+M)T_{\mathrm{f}})\nonumber\\
&&\times \left(\!\rho(q,\kappa_i)\!-\!1\!\right)e^{-j 2 \pi \frac{\left[k-k_{i}+q\right]_N}{N}}\widetilde{X}\!\left[\left[k-k_i+q\right]_N,\left[l-l_i\right]_M\right].\nonumber\\ \label{ysf}
\end{IEEEeqnarray}
Therefore, by combing \eqref{ycf} and \eqref{ysf}, the DD-domain input-output relationship can be expressed as \eqref{DD1}. This completes the proof.
\bibliographystyle{IEEEtran}
\bibliography{IEEEabrv,ref}

\end{document}